\documentclass[preprint]{aastex}

\def\etal{et~al.}

\shorttitle{Globular Clusters in Coma Ellipticals}
\shortauthors{Harris et al.}

\begin{document}

\title{The Globular Cluster Systems in the Coma Ellipticals. IV:  
WFPC2 Photometry for Five Giant Ellipticals\footnote{Based on 
observations with the NASA/ESA {\sl Hubble Space Telescope}, 
obtained at the Space Telescope Science
Institute, which is operated by the Association of Universities for
Research in Astronomy, Inc., under NASA contract NAS 5-26555.}
\footnote{This research used the facilities of the Canadian Astronomy 
Data Centre operated by the National Research Council of Canada with 
the support of the Canadian Space Agency.}
 }

\author{William E.~Harris}
\affil{Department of Physics and Astronomy, McMaster University, 
Hamilton ON L8S 4M1, Canada }
\email{harris@physics.mcmaster.ca}

\author{J.~J.~Kavelaars}
\affil{Herzberg Institute of Astrophysics, National Research Council, 5017 West Saanich Road, 
Victoria BC V9E 2E7}
\email{jj.kavelaars@nrc-cnrc.gc.ca}

\author{David A. Hanes}
\affil{Department of Physics, Queen's University, Kingston ON K7L 3N6, Canada}
\email{hanes@astro.queensu.ca}

\author{Christopher J. Pritchet}
\affil{Department of Physics and Astronomy, University of Victoria, Box 3055, Victoria, BC V8W 3P6, Canada}
\email{pritchet@phys.uvic.ca}

\author{W.~A.~Baum}
\affil{Department of Astronomy, University of Washington, Box 351580, Seattle WA 98195}
\email{baum@astro.washington.edu}

\begin{abstract}
We analyze photometric data in $V$ and $I$ 
for the globular cluster (GC) systems in five of the
giant ellipticals in the Coma Cluster:  NGC 4874, 4881, 4889, 4926, and
IC 4051. All of the raw data, from the Hubble Space Telescope WFPC2
Archive, are analyzed in a homogeneous way so that their five 
cluster systems can be strictly intercompared.  
We find that the GC luminosity functions 
are quite similar to one another and reinforce the common nature
of the mass distribution of old, massive star clusters in gE galaxies.  
The GCLF turnover derived from a composite sample of more than 
$9,000$ GCs appears at $V = 27.71 \pm 0.07$ ($M_V = -7.3$), and
our data reach about half a magnitude fainter than the turnover.
We find that both a simple Gaussian curve
and an evolved Schechter function fit the bright half of the 
GCLF equally well, though the Coma GCLF is broader and has a higher
``cutoff mass'' ($M_c \sim 3 \times 10^6 M_{\odot}$) than in any
of the Virgo giants.  These five Coma members
exhibit a huge range in GC specific frequency, from a low
of $S_N \simeq 0.7$ for NGC 4881 up to $\simeq 12$ for IC 4051
and NGC 4874.
No single formation scenario appears able to account for these
differences in otherwise-similar galaxies and may require
carefully prescribed differences in their merger history, gas-free
versus gas-rich progenitors, GC formation efficiency, initial
density of environment, or tidal
harassment within the Coma potential well.
The supergiant cD galaxy NGC 4874 has the richest 
globular cluster system known, probably holding more
than 30,000 clusters; its true extent is not yet determined
and may extend well out into the Coma potential well.
For the three biggest GC systems (NGC 4874, 4889, IC 4051),
analysis of the $(V-I)$ color distributions shows that all
three populations are dominated by red, metal-rich clusters.
Their metallicity distributions also may all have the normal
bimodal form, with the two sequences at mean colors
$\langle V-I \rangle$(blue) $\simeq 0.98$ and $\langle V-I \rangle$(red)
$\simeq 1.15$.  These values fall along the previously established
correlations of mean color with galaxy luminosity.
However, the color distributions and relative numbers of metal-rich
clusters show intriguing counterexamples to a trend established
by Peng et al.~2008 (ApJ 681, 197) for the Virgo galaxies.
For the brightest Virgo ellipticals, they find that the red GCs
make up only $\sim 30$\% of the cluster population, whereas in our
similarly luminous Coma galaxies they make up more than half. 
At the very highest-density and most massive regimes represented
by the Coma supergiants, formation of metal-rich clusters seems to
have been especially favored.
\end{abstract}

\keywords{Galaxies:  Formation -- Galaxies:
Individual -- Galaxies:  Star Clusters}

\section{Introduction}

The Coma cluster is the nearest example of an Abell cluster which holds
a truly rich collection of E/S0 galaxies in a dynamically evolved
environment.  As such, it provides a unique locale for the study of
globular cluster systems (GCSs) under higher-density and
dynamically ``older'' environmental conditions
than in Virgo, Fornax, and smaller nearby groups of galaxies
\citep{har01}.

The first HST-based photometry of a GCS in Coma was for 
NGC 4881 by \cite{bau95}, a giant elliptical in the cluster core.
In Papers I and II of our series on the Coma cluster
\citep{kav00,har00}, we discussed the GCS in
NGC 4874, the supergiant cD-type elliptical that lies near the center of
the larger Coma potential well.  In \citet{bau97} and 
in our Paper III \citep{woo00}, 
similar material was presented for IC 4051, another giant 
on the outskirts of the Coma cluster core region.   In the present
paper, we discuss new results (GCS radial distribution, total population,
metallicity distribution, and luminosity function) for two additional
Coma members, NGC 4889 and 4926, that have not previously
been published.  Finally, we reanalyze the WFPC2 Archive 
data for NGC 4874, 4881, and IC4051 to ensure that all of it is on
a homogeneous photometric basis, and discuss 
all five galaxies together.

The three Coma members that we studied earlier through HST
(NGC 4874, 4881, IC 4051) already present a striking range of GCS properties.
NGC 4881 has an extremely low-specific-frequency system that is quite surprising
for a giant E galaxy in such a rich environment.
By contrast, IC 4051 has a 
system in the high-$S_N$ range that would normally be associated
with central supergiant
cD galaxies \citep{har01}, but one that is also spatially quite compact,
perhaps due to severe tidal truncation from the Coma potential.
NGC 4874, the central supergiant, has a high-$S_N$ 
system with a very large radial extent, perhaps spanning the
entire Coma core.  These systems hint that we may not yet have seen the entire
variety of GCS characteristics that this environment has to show.

The globular cluster populations in the Coma galaxies have
also attracted attention through ground-based imaging, although
their $\sim$100-Mpc distance has always made such efforts extremely
challenging.  \citet{har87} and \cite{tho87} first obtained deep
enough imaging to reveal 
the GCs in NGC 4874 with cameras on the CFHT, though these studies were
useful for no more than the roughest estimates of specific
frequency (that is, the number of GCs per unit galaxy luminosity).
\citet{bla95} used a combination of surface brightness fluctuation 
(SBF) techniques and resolved-GC photometry
to estimate the GC populations in NGC 4874 and 4889.
More recently, \citet{mar02} used  
SBF to measure the cluster populations in 17 Coma galaxies.  
From this survey, they confirmed
the rather bewildering variety of specific frequencies from
galaxy to galaxy -- from a cD-like high of $S_N \simeq 12$ down
to the $S_N \sim 1$ level that we normally associate with 
much more GC-poor spiral and dwarf galaxies.  Furthermore,
from their entire sample of 17 galaxies
there is no trend of $S_N$ with either galaxy luminosity
or location (radius from Coma center).  However, by restricting
their sample to only their five galaxies within ``subgroup 1'' of
\citet{gur01}, which is the subcluster centered on NGC 4874 itself,
they conclude that $S_N$ may indeed be correlated with environment
in the sense that $S_N$ decreases with increasing distance
from the group center.  A similar trend had already been 
proposed by \citet{bla97} from a sample of specific frequencies
for 23 galaxies in 19 rich clusters.  This trend is also reminiscent
of the new discussion by \citet{pen08} for the Virgo galaxies, in which 
they show that the dE galaxies in Virgo have higher $S_N$
values closer to the central giant M87.  

Although SBF photometry has proven to be effective in measuring
the total GC populations in E galaxies, it cannot say much about
other essential characteristics such as the GC color 
or luminosity functions, which are the key visible tracers
of the GC metallicity and mass distributions.  For these,
we need the deep photometry of individual GCs that the HST
cameras can provide.  In Coma, there are five such galaxies
with available WFPC2 imaging data that we can discuss within
a strictly homogeneous photometric system.  These are summarized
in Table \ref{tab:targets}, where we list the
galaxy luminosity (with total magnitudes adopted
from RC3), heliocentric radial velocity, and
projected distance from the center of Coma (assumed to be
at NGC 4874).  For comparison, the core radius of the entire Coma galaxy
distribution is $r_c \simeq 15'-20'$ (see Paper II) and
its overall mean heliocentric velocity is 6917 km s$^{-1}$ \citep{col96}.  

Much previous work in the literature can be found on the
spatial and velocity substructures within Coma.  To first
order, identifiable subclusters can be found centered on
the supergiant cD galaxy NGC 4874 (the center of the
biggest subsystem), the other outlying cD NGC 4839, and
the supergiant NGC 4889 \citep[for much more detailed 
discussions, see][among others]{fit87,bai90,mer94,col96,gur01}.
These subgroupings also appear in the intracluster X-ray gas
\citep{neu03,ada05}.  NGC 4889 has a total luminosity comparable with NGC 4874, but
is structurally more compact than NGC 4874, without a
cD-type envelope.  The other three systems in our WFPC2 study 
(NGC 4881, 4926, IC 4051) are more normal
giant ellipticals and are all among the ten brightest Coma members.
Only NGC 4926 is clearly outside the main cluster core.

In the following sections, we describe the raw data used for our
program, the subsequent analysis, and lastly a comparison of the five
galaxies.   We assume here a distance modulus for Coma
of $(m-M)_0 = 34.97 \pm 0.13$ or $d=98.6 \pm 6.1$ Mpc. 
This distance results from our adopted redshift of
$7100 \pm 200$ km s$^{-1}$ \citep{col96,kav00} and
a Hubble constant $h = 0.72 \pm 0.04$ \citep{fre01,spe07}.
At this distance, 10 arcseconds on the sky is equivalent to 
a linear scale of 4.8 kpc.

\section{The Database}

The raw WFPC2 data for these galaxies were all long exposures drawn from the
HST Archive.  The images were originally acquired as parts of
programs GO-5233, 6283 (NGC 4881 and IC 4051; Westphal PI) and GO-6104, 8200 
(NGC 4874, 4889, 4926; Harris PI).
Total exposure times and filters are summarized in Table \ref{tab:data}.
The raw capabilities of the WFPC2 camera are now exceeded by the
newer ACS and WFC3, which have wider fields of view and higher
sensitivity particularly in the blue.  Eventually, it will be possible
(for example) to study the color/metallicity distributions and
spatial distributions of the Coma galaxy GCs with higher precision
with these better cameras \citep{car08}.  However,
the WFPC2 images for four of these Coma targets 
already reach quite far into their globular cluster luminosity
functions (GCLFs) and, to this point, make up the deepest available sample of 
Coma GC material to work with.  For this reason, we believe it is 
worth discussing the combined material in a homogeneous way. 
We will concentrate on two main questions:  first, are their GCLFs
the ``standard'' Gaussian-like ones seen in other large E galaxies?
And second, how similar are their color distributions to one another
and to the conventional bimodal form found in
other large galaxies, as far as we can gauge them within the 
limitations of the $(V-I)$ color index?

Our data reduction 
follows the same procedures outlined in detail in Papers I-III.
Master $V$ and $I$ frames free of cosmic rays and bad-pixel artifacts
were generated by registering and median-combining the individual 
WFPC2 exposures.  When combining the exposure sequences,
we treated each of the four CCD chips separately (we did not rely on
the Archive WFPC2 Associations, performing the image combining operations
ourselves from the raw exposures).  For all five fields, the PC1 chip
is centered on the galaxy.  To facilitate the detection of faint starlike
objects (which include the globular clusters we are looking for; see below),
on the combined PC1 images we generated a smooth model
of the galaxy isophotes with the {\sl STSDAS/ELLIPSE} package
and subtracted it from the image. In the WF2,3,4 frames we median-filtered
the image to remove the more gradual isophotal gradient from the outer parts of
the galaxy.  These steps were done iteratively along with 
the {\sl DAOPHOT} star-finding and removal
to produce the cleanest possible master images for final stellar photometry.

At the 100-Mpc distance of Coma, a typical globular cluster half-light diameter
of $\simeq 6$ pc subtends $0.012''$, an order of magnitude smaller
than the resolution of the telescope.  This feature of the data
makes the GCs indistinguishable from stars in 
appearance, and carries two major advantages for
the photometry:  first, it is straightforward to carry out normal PSF-style
object detection and photometry on these images; and second, 
most of the background population of small, faint galaxies 
can be easily distinguished and removed from the sample, 
minimizing the field contamination.

\begin{figure}
\plotone{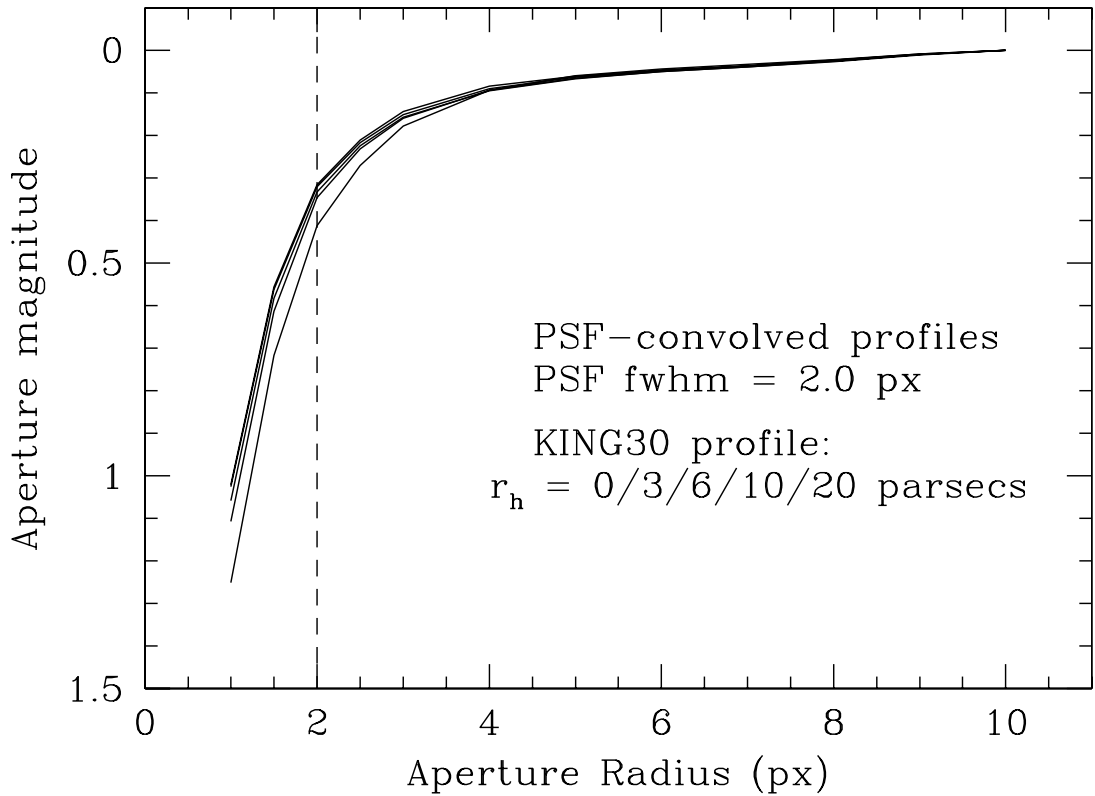}
\caption{Aperture growth curves for various synthetic globular clusters.
Each curve shows the results for a KING30 model cluster with a
particular half-light radius $r_h$, placed at the 100-Mpc distance
of Coma and then convolved with the WFPC2 point spread function.
The magnitude within aperture radius $r(ap)$ is shown for five model clusters
with $r_h = 0, 3, 6, 10$, and 20 parsecs, a range which generously
brackets real globular clusters.  The curves are plotted
to osculate at $r=10$ px; the top curve is for
$r_h = 0$ (that is, a starlike object), while the bottom curve is for
$r_h = 20$ pc, which would be a very
extended object resembling a UCD; see text.
The curves for $r_h = 0$ and 3 pc are indistinguishable.
For typical globular cluster radii $r_h < 6$ pc, the PSF-convolved
profiles can be treated as stars at this distance, and even
for UCD-type objects the profiles are starlike for $r > 5$ px.
\label{fig:apcor}}
\end{figure}

In Figure \ref{fig:apcor} we show the results of some simulations
that demonstrate the validity of the PSF-fitting procedure for
this material.  Here, we have used the tools in the {\sl baolab/ishape} code
of \citet{lar99} to generate simulated cluster profiles of
various half-light radii $r_h$ in a realistic range; project
them to the Coma distance; convolve them with the WFPC2 point spread
function; and finally place them into an image for remeasurement.
For these tests we specifically used our empirical
point spread function derived for
the NGC 4926 WF fields, but any of the other targets would have done equally well.
As is extensively discussed by \citet{lar99}, a PSF-convolved cluster
profile will yield a measurable and systematically correct $r_h$
with the use of fitting codes like {\sl ishape}
if the $FWHM$ of the intrinsic cluster profile
is larger than about 10 percent of the $FWHM$ of the PSF itself.  
Objects smaller than this are effectively indistinguishable from stars.

Fig.~\ref{fig:apcor} shows an equivalent way to express this result.
We generate PSF-convolved profiles for model clusters of different
radii:  $r_h = 0, 3, 6, 10, 20$ parsecs.  The smallest $r_h$ 
obviously corresponds
to a pure stellar profile; $r_h = 3$ pc corresponds to an average
Milky Way globular cluster \citep[see the data in][]{har96}; $r_h = 6$
pc is at the upper end of the normal GC distribution; and $r_h = 10$ and
20 pc correspond to profiles for more extended objects resembling
dE nuclei or Ultra-Compact Dwarfs (UCDs); see, for example, \citet{evs08}.
For the template GC profiles we used the KING30 model in {\sl ishape},
which closely matches the mean central concentration of real GCs
\citep{har96,lar99}.  In Fig.~\ref{fig:apcor}, the curves of growth
(aperture magnitude enclosed within radius $r$ in pixels) are plotted
for each of the five models.  At the measurement radius of 2 px that
we adopted for all our photometry, we then read off the 
magnitude {\sl difference} $\Delta m$ between the pure stellar curve ($r_h = 0$)
and the curve for the more extended object.  This difference gives
us the magnitude correction that we would have to apply to the
actual, PSF-convolved GC profile in order to put its measured
magnitude onto the same scale as
a pure PSF.  In principle, we could then use {\sl ishape}
to solve for all the individual $r_h$ values and then use this
grid of curves to read off the necessary magnitude correction
to apply to each one.

In practice, what we find is that the curve for $r_h = 3$ pc
(that is, a ``baseline normal'' GC) is indistinguishable from 
that of a star.  For any $r_h < 6$ pc, 
the correction is $\Delta m < 0.01$ mag and thus negligible
compared with the internal scatter of the photometry (see below).  
It is only for UCD-like systems
at $r_h \sim 20$ pc where the correction approaches $\sim 0.1$ mag.
As further confirmation, we note the work of \cite{weh07} for the
NGC 3311 globular cluster system, which is at roughly
half the Coma distance.  Even there, only a handful of the
largest, brightest, UCD-like clusters could be seen to be marginally
resolved on WFPC2 images.

The object-finding and photometry was carried out with the 
normal {\it DAOPHOT/ALLSTAR} codes
\citep{ste94}. In each field, in each of the $V$ and $I$ filters,
and on each one of the four camera quadrants (PC1, WF2,3,4 independently),
we generated empirical point spread functions (PSFs) from moderately
bright isolated stars.  The instrumental magnitudes returned by
these codes were CTE-corrected, then transformed to the standard $(V,I)$ system 
\citep{hol95} after aperture correction to the nominal
$0.5''$ radius specified by the transformations.  

A problem encountered at this stage was that 
the fields of some of our galaxies (NGC 4926 and 4881 particularly)
are so ``clean'' of foreground
stars -- more so than in the other Coma fields --
that the PSFs, in the end, relied on fewer and fainter stars 
than we would have preferred, particularly in the $I-$band frames
which had shorter total exposure and thus lower signal-to-noise
than the $V$ frames.  The main step which might be affected
adversely is the aperture correction (mean difference between the 
PSF-fitted instrumental magnitudes and the $0\farcs5$ whole-aperture
magnitudes) and thus the zeropoint of the final $I$ photometric scale.
Although the $V$ magnitude scale should be correct to $\pm0.02$ mag, the
scale in $I$ may still be uncertain chip-to-chip
to no better than $\pm0.05$ mag, a worry which
is borne out to some extent in the dispersion of the cluster colors 
once we combine the data from all four chips (see below).

A high proportion of the background contamination (mostly faint,
slightly nonstellar background galaxies)
can be eliminated objectively, first through the use
of radial-moment image analysis and second through rejection of any
objects not found in both filters (extreme red or blue colors).
As in our previous papers, for the radial moments we use the $r_1$
moment as implemented in \citet{har91} along with artificial-star tests to
define the boundaries of the stellar sequence (see Papers I and III for
complete discussion).  

\begin{figure}
\plotone{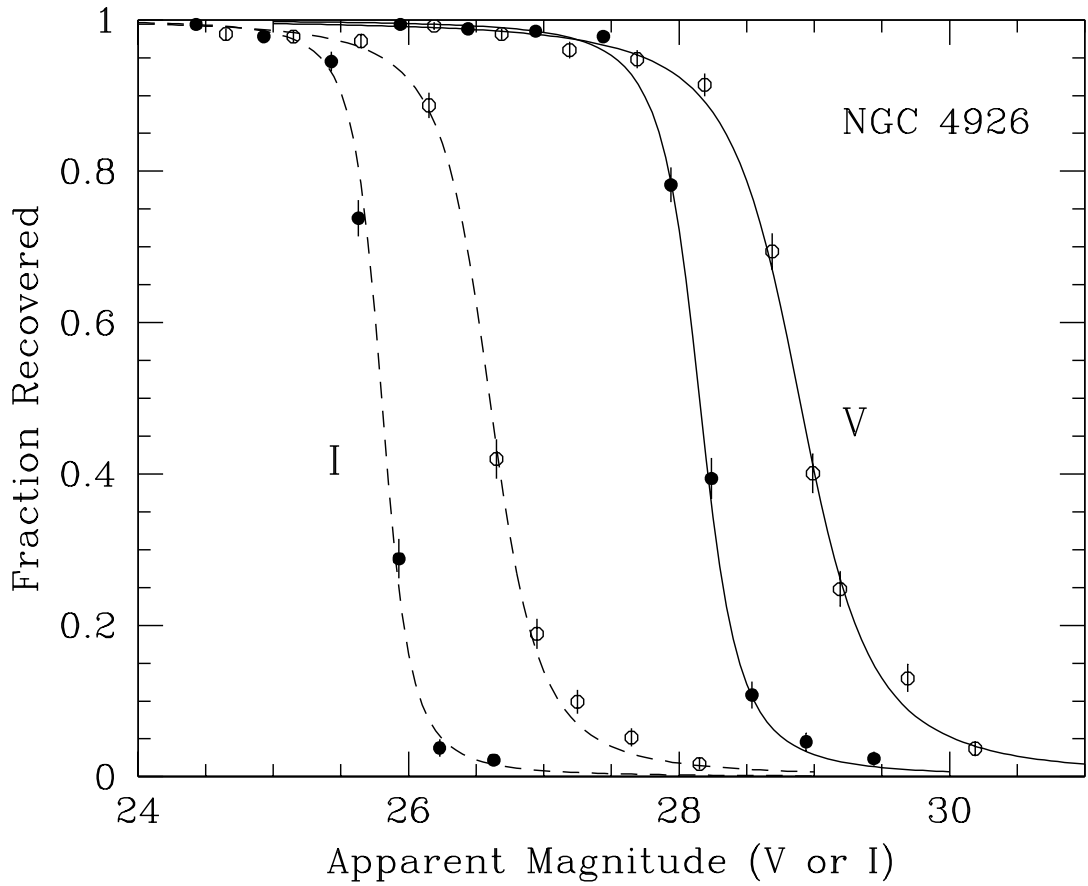}
\caption{Completeness of detection of the photometry for the NGC 4926 field.
The smooth curves are the interpolation functions given in the text:
the solid lines represent the mean for the $V$ frames while the
dashed lines represent the $I$ frames.  In each pair the left-hand
curve (solid dots) shows the curve for the PC1 CCD chip, 
while the right-hand curve (open circles) shows the WF CCD chips.
\label{fig:completeness}}
\end{figure}

Lastly, artificial-star simulations of the same type used in our previous
work were used to define the completeness of detection $f(V,I)$, an
important issue particularly for determining the GC luminosity function.  
Even in the innermost regions around each galaxy, the photometry is
not affected by crowding, so the completeness and internal measurement
scatter are determined only by the faintness of each object and
the surface brightness of its local background.  Averaged
results over many simulated runs (where we add several hundred 
artificial stars to the images, then re-reduce them through the 
same {\sl DAOPHOT} measurement sequence) were used to define 
the $f-$distributions for each field and each chip.
One sample of these is shown in Figure \ref{fig:completeness}.  
Approximate fits
to the trend of $f$ with magnitude are given by modified versions of the
interpolation functions used in our previous papers,
\begin{equation}
f = 0.5 \Big( 1 - {{\alpha (m-m_0)} \over {\sqrt{1 + \alpha^2 (m-m_0)^2}}}\Big) 
\end{equation}
The two free parameters in this function are the ``completeness limit''
$m_0$ at which $f$ reaches 50\%; and the slope $\alpha$, which governs
how steeply $f$ declines as it passes through the $m_0$ midpoint
(see Fig.~\ref{fig:completeness}).
The various values of $(m_0, \alpha)$ that we found from the artificial-star
tests are given in Table \ref{tab:completeness}.
For the same field, $m_0$ is noticeably different on the PC1 chip
than on the three WF chips in the sense that PC1 does not reach as faint.
This difference is primarily due to the much higher background light
on PC1.  We found no dependence of the limiting magnitude on galactocentric
radius for the WF chips.  For the PC1 fields, $m_0$ becomes noticeably
brighter within $\simeq 5''$ of the galaxy centers, but as will be seen below,
we made little or no use of the data in this innermost region, and none
of our conclusions depend significantly on it.

The artificial-star tests  
showed that the random measurement uncertainties are
reasonably well described by a simple interpolation function
$e(m) \simeq 0.03 + a \cdot {\rm exp}(V-b)$, for constants $a,b$.
For NGC 4874, 4889, and IC 4051 we find $a = 0.04, b_V = 26.2, b_I = 25.3$:
for NGC 4926 $a = 0.01, b_V = 25.2, b_I = 23.4$; and for NGC 4881
$a = 0.04, b_V = 24.5, b_I = 24.5$.
These approximations also match well with the internal photometric
uncertainties returned by the {\sl ALLSTAR} solutions themselves, 
consistent with cases such as this where crowding is quite unimportant
and the measurement uncertainties are dominated simply by the object
faintness and the local background light.

The artificial-star tests and plots of the raw data versus radius showed
that within $r \simeq 4''$ (40 pixels, or 2 kpc in linear distance) 
of the galaxy centers the data 
become severely incomplete because of the very high background light.
The data within this innermost region around each galaxy were 
simply eliminated from our discussion.

\begin{figure}
\plotone{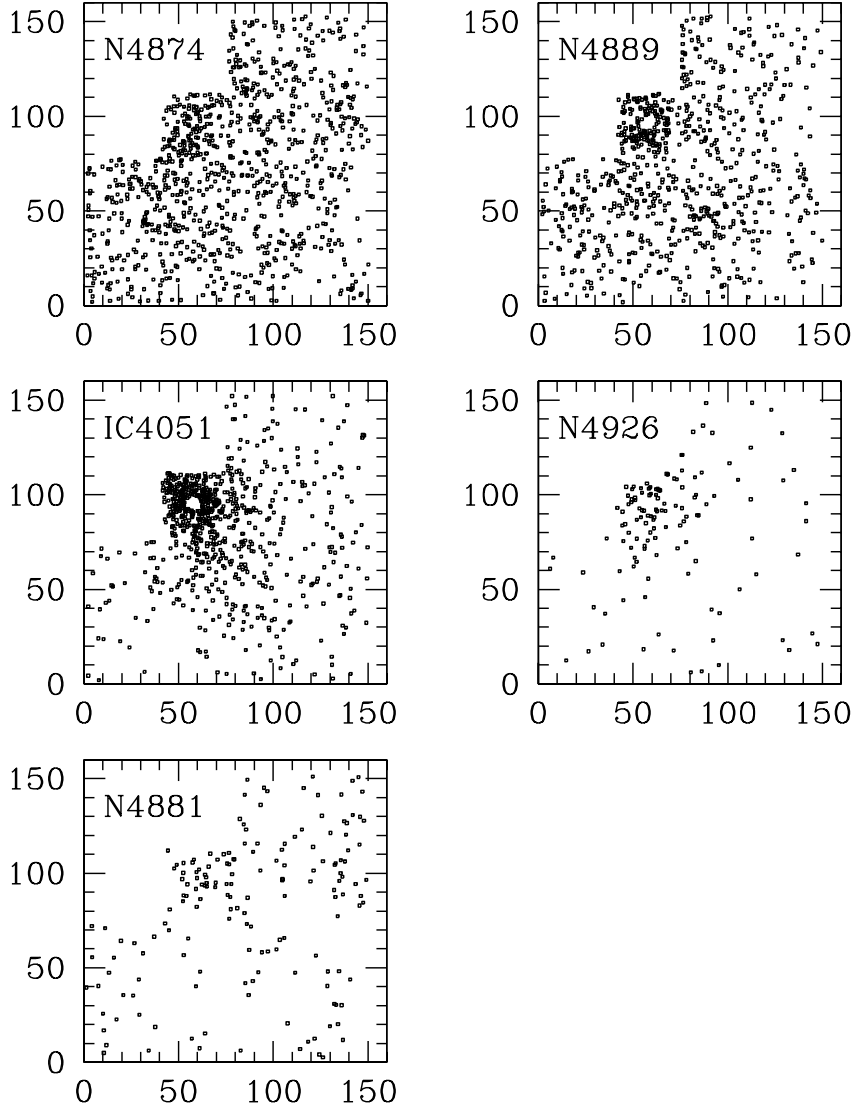}
\caption{Distribution of detected starlike objects around each
of the five target galaxies in the WFPC2 field.  In each case
the objects brighter than $V = 26.0$ are shown, with no
additional selections by color.  The scale shown is in arcseconds.
The orientation is such that the PC1 chip is at upper left, 
WF2 at lower left, WF3 at lower right, and WF4 at upper right.  
In every case the center of the galaxy is nearly at the center
of the PC1 chip.
\label{fig:xy}}
\end{figure}

In Figure \ref{fig:xy} we show the spatial pattern of the brighter
``starlike'' objects ($V < 26$ and detected in both $V$ and $I$) 
around each of the five galaxies.
As will be seen below, the magnitude range $V < 26$ includes the brighter
portion of the GC luminosity function and includes only a small
proportion of field contamination, so these graphs give a
useful visual impression of the size and spatial extent of
the GC distributions.  A striking variety of spatial structures 
is already evident.

The raw color-magnitude distributions (CMDs) for the GCs are shown in
Figure \ref{fig:cmd5}.  The large differences in total GC populations
show up here as well, but it is also evident from the CMDs that
the {\sl color} distributions in $(V-I)$ are rather similar.
We will quantify these features in the following sections, as well
as the luminosity functions.

\section{The Spatial Distributions}

Assessing the intrinsic spatial distributions of the globular
cluster systems around these galaxies, as well as their
luminosity functions and color distributions, first requires
knowing the level of background contamination.  For NGC 4874
and 4889, we can estimate only upper limits to the background
level (i.e., the number of field contaminating objects per unit
area) from those images alone, because the GCS probably extends
well beyond the borders of the WFPC2 images.  Fortunately,
for IC 4051 and NGC 4926 the story is different.  These two
GC systems are compact enough that the measured numbers of
objects drop off to near-constant levels for radii beyond
$\simeq 50''$ (equivalent to 25 kpc).
Since the exposure depths for
these four fields were very similar, and they are all located
in the high-latitude Coma region, we therefore decided to use the outer
parts of the NGC 4926 and IC 4051 fields to define an average ``global''
background distribution in magnitude and color, which could
then be used for the three galaxies that had by far
the most populous GCSs (namely, NGC 4874, 4889, and IC 4051 itself).  
The raw number-density plots for
all the individual galaxies are shown in Figure \ref{fig:rad4}.
This plot shows the radial falloff of the GC population for the
magnitude range $22.0 < I < 25.5$ over which the photometry is
essentially 100\% complete.
Experiments with the IC 4051 and NGC 4926 fields
led us to adopt the ``global background field'' as
consisting of all measured objects farther than  $80''$ from
the center of IC 4051 and farther than $70''$ from the center
of NGC 4926.  The total area of this composite background region is 
3.98 arcmin$^2$.

For NGC 4881 and 4926, which have GCSs that are both compact {\sl and}
sparsely populated, we used the outer regions of only their own
fields to define a local background, preferring in this case
not to average in data from other fields that might have been
subtly different in measurement selection and intrinsic
background population (any such minor differences
are far less important for the enormously more populous
GCSs in the other three).
The limiting magnitudes of the NGC 4881 exposures
are also shallower than for the other four and so both the GC numbers
and the background are especially low for it.

The spatial distributions, {\sl after subtraction of background}, are
shown in Figure \ref{fig:lograd}.  Unless otherwise noted below,
the GC counts are for $V < 27$, well within the range of
high completeness.
Here, the distributions are shown in
the classic E-galaxy form of log $\sigma$ versus $r^{1/4}$.
In each graph we superimpose the available surface-brightness
photometry for each galaxy, in order 
to compare the GCS spatial structure with that of the integrated halo
light of the parent galaxy.  
Individual comments are summarized below.

\noindent {\sl NGC 4874:}  The integrated light profiles are
from \citet{jor92} in $r$ (dashed line) , and \citet{pel90} in $R$
(solid line).  
At radii $r \gtrsim 5'$, 
the surface brightness profile begins to overlap that of
NGC 4889 and for both galaxies becomes much more difficult to trace further outward.
Its GC system is clearly very extended, going well beyond the
WFPC2 field boundaries (Paper II and Fig.~\ref{fig:xy}).
Despite these restrictions, it is already obvious that
the GCS has a distinctly shallower 
profile than the halo isophotes.  More detailed discussion of
the radial profile can be found in \citet{har00} and our conclusions
here are the same.  The primary difference in the current data is
that we were able to use a global background field with 
strictly homogeneous photometry and twice
the effective area.  As we discussed in Paper II,
an important point of emphasis peculiar to NGC 4874 is the
long-standing question whether cD-type centrally dominant
galaxies may have accreted a high fraction of their GCs from
their intracluster environment.  Notably,
its GCS is quite a bit more centrally concentrated
than is the Coma X-ray gas, which in the region close to NGC 4874
has a near-constant surface brightness (see Fig.~3 of Paper II
for the specific comparison).  If there is a GC population component
with a similarly near-flat spatial distribution in the Coma core
region, it clearly does not show up in any important way
within $\sim 100$ kpc of the galaxy.
For these reasons it is possible to argue
that the GCS predominantly belongs to the central galaxy rather than to
the Coma potential well at large.  

\noindent {\sl NGC 4889:}  
The surface brightness profiles are
from \citet{jor92} (dashed line) and \citet{pel90} (solid line).
For this galaxy, the results fall
into a familiar pattern whereby the GCS is moderately more
extended (shallower) than the halo light throughout the entire
run of our data, though both have steeper profiles than their counterparts
in NGC 4874.  Combined with the moderate specific frequency
and high metallicity distribution that we find for NGC 4889 (see below),
we interpret it as a normal, though extremely high luminosity,
elliptical.  

For both NGC 4889 and 4874, much wider-field 
imaging with similar depth will be needed to gauge properly how
extended their cluster systems are at larger radius, since
for $r \gtrsim 5'$ they overlap each other rather heavily.
The hints from the current data are that the NGC 4874 system
probably dominates more and more at larger radii because of
its shallower profile, but this remains to be seen.
Wider-field data are also the only way to assess the true contribution,
if any, of intracluster GCs within the Coma core.
A roughly similar case has been studied in the Hydra cluster
\citep{weh08}, where two giants NGC 3309 and 3311 sit close
to each other in the cluster core and one of them (NGC 3311)
is a central cD.  With the right data, a numerical
solution can be performed to derive the two total GC populations
and spatial distributions simultaneously.  For Hydra, the result
shows that the cD NGC 3311 is dominant and NGC 3309 unimportant; but in Coma, 
both the central giants have very large cluster populations and
the actual degree of overlap is not yet known.

\noindent {\sl IC 4051:}  
The surface brightness profiles are from \citet{jor92} (solid line)
and \cite{str78} (dashed line), both in $R$.
The high radial velocity of IC 4051 suggests that it is now
plunging through the Coma core \citep[see][]{woo00}.
This galaxy clearly has a very populous GCS, but it has a
spatial concentration remarkably higher than either NGC 4874 or 4889.
A more extensive discussion of the comparison between halo light
and GCS profile is in Paper III; we conclude similarly that
the GCS is slightly shallower for $r \lesssim 15''$ but has
a steep outer cutoff past $80''$ that may follow the halo light.

\noindent {\sl NGC 4926:}
The surface brightness profiles are from \citet{jor92} (dashed line)
and \citet{jan00} (solid line), both in $R$. 
The GC surface density is far lower than in IC 4051, but both it
and the halo light have a similarly steep falloff.
We see no strong evidence that the GCS and halo light differ
in structure.
As we suggested in Paper III, this relatively compact structure and
steep outer cutoff may be due to tidal trimming (``harrassment'') from
the Coma potential well \citep[e.g.,][]{moo96}, since it is plausible
that both IC 4051 and NGC 4926 have oscillated through the Coma core several times
during their lifetimes.

\noindent {\sl NGC 4881:}
The surface brightness profile is from \citet{tho87} in $\lambda523$nm.
Here we used our data for $V < 26.5$, slightly fainter than the
normal completeness limit, but we were forced to do this to obtain
any kind of estimate of the radial falloff of the GCS. 
Within the large uncertainties of the GC counts, there is no
difference between the GCS distribution and the halo light.

The numerical data for the measured objects as shown in Fig.~\ref{fig:lograd}
are summarized in Table \ref{tab:rad}, in which
column (1) gives the mean annulus radius $\langle r \rangle = 
\sqrt{r_{inner} \cdot r_{outer}}$;
column (2) the number of objects lying in the annulus before
background subtraction,
column (3) the annular area in arcsec$^2$ enclosed within the WFPC2 field boundary;
and column (4) the number of objects per unit area before background
subtraction.

\section{Luminosity Functions}

For NGC 4874, 4889, 4926, and IC 4051, the original exposures
in $V$ (though not $I$) were deliberately designed to 
trace the GCLF down to levels as faint
as the expected GCLF ``turnover point'' (peak frequency in number
per unit magnitude).  We therefore used the $V$ data alone to 
measure the luminosity functions.  A minor loss in using only
one band is the inability to reject any candidate objects by
color, and although the relative field contamination is quite
low at the brighter magnitudes (Fig.~\ref{fig:cmd5}) compared with the
GCs, it increases toward the faint end.

The results for the four galaxies with deepest photometry
(i.e. all except NGC 4881) are summarized in Table \ref{tab:lf}
and Figure \ref{fig:lf}, which show the number of deduced
GCs per quarter-magnitude bin, fully corrected for 
photometric incompleteness and background-subtracted.  
For NGC 4874 and 4889, we took all measured objects
from $5''$ to $120''$ from the center of the galaxy
to define the GC signal,
and used the global background defined above, normalized
to the same equivalent area, to subtract off field
contamination.  
For the much more compact IC 4051 system, 
we used the region from $5''$ to $80''$ 
and subtracted off the global background.
Finally, for NGC 4926 we took the GC region to be
$5'' - 70''$ and again used the global background.
In Table \ref{tab:lf} we also give the LF of
the global background population and 
the total areas of each region are listed in the last line.

If the GCLFs of the Coma galaxies resemble those found
in other giant ellipticals \citep{har01,wat06}, we would expect them to
have a Gaussian-like distribution in number per
unit magnitude, with a peak frequency (turnover point)
roughly at $V^{to} \simeq 27.7$, and a dispersion $\sigma_V \simeq 1.4$.
This question is particularly relevant for the two supergiants,
because it gives us the opportunity to extend tests for any systematic
change of the GC mass distribution function 
upward to bigger galaxies than we are able to access 
in Virgo or Fornax.  For example, if the supergiants are
simply merger products of smaller galaxies that formed all their
GCs beforehand, then their product GCLFs should be roughly
similar to those in smaller galaxies now.  If differences appear,
then it may point in turn to a different formation process.

The expected turnover point is within reach of our data,
which means that the entire bright half of the GCLF should
be visible.  For $V \lesssim 27.5$ the photometry in all four galaxy
fields is nearly 100\% complete,
and comparisons of the GCLFs showed that they are
virtually identical in shape over that range, independent of
any particular fitting function.  An individual puzzling result
is for the faint end of the NGC 4889 LF, which according to our raw numbers
seems to keep rising past the expected turnover point, unlike
the other three systems.  We have no definite explanation for this
except to note that the anomaly occurs just near the photometric
completeness limit, where a number of different difficulties with
the combined background and completeness corrections can come into play.
Conservatively, we have chosen to use only the data 
for $V < 27.75$ for both NGC 4874 and 4889.

Fainter than the expected turnover, 
the data in the four fields start to differ quite strongly in their
completeness functions, and simply adding them up at each
magnitude bin would no longer be valid.
To make a reasonable estimate of the 
GCLF shape to slightly deeper levels, we therefore constructed a composite LF
by the following  procedure, which ensures that we use only
the data where the completeness $f_V$ is higher than 50\%
in each bin:

\noindent (a) For $V < 27.75$, the composite LF is the direct
sum of all four fields, as listed in Table \ref{tab:lf}, with the
individual completeness corrections applied.  Over this magnitude range,
$f_V > 0.5$ for all four fields.

\noindent (b) For $27.75 < V < 28.25$,
the LF is the direct sum
of the NGC 4926 and IC 4051 fields, scaled up by a factor of 3.287
to allow for the fact that these two contribute 30.4\% of the total
for $V < 27.75$.

\noindent (c) For $28.25 < V < 28.50$, we drop IC 4051 and use
only NGC 4926, scaled up by a factor 22.585.  

The result is shown in Figure \ref{fig:lfsum}.  The explicit assumption
we use is that the GCLFs in these four galaxies are intrinsically
similar enough to permit this scaling procedure to work.  We can
test that assumption only over the bright half, but the reason for
doing it is to gain the highest statistical confidence we can in
carrying the LF past the turnover point.  As is also obvious from
the scaling procedure just described, the internal uncertainties
grow rapidly toward the faint end because of all three factors
of increasing field contamination, photometric incompleteness,
and scaling factor.  If the raw number of objects in
the bin is $N$ over area $A$, the number of background objects 
is $N_b$ over area $A_b$, the
mean photometric completeness factor is $f$, and the population
scaling factor is $s$, then the fully corrected total in the bin
is $N_{GC} = (N-N_b A/A_b) s / f$ and its uncertainty due to
sampling statistics alone is 
$\pm (N+N_b (A / A_b)^2)^{1/2} s / f$. 
In practice, this can be treated as only a lower limit
to the true uncertainty, since the background may have non-Poissonian
fluctuations over scale lengths of a few arcminutes.

Next we want to use the combined GCLF to make the best estimate
of the turnover magnitude and the intrinsic width or dispersion
of the whole distribution.
Extensive discussions of interpolating model fits to the 
GCLF are given in Papers I and III and we will not repeat them
at length here.  We start by fitting the classic Gaussian function
to the binned data in Fig.~\ref{fig:lfsum}, solving simultaneously
for the turnover magnitude $V^{to}$ and dispersion $\sigma_V$,
by $\chi^2$ minimization.  Calculations in a finely spaced grid over a wide range in both
parameters shows a well behaved global minimum 
at $V^{to} = 27.71 \pm 0.07$, $\sigma_V = 1.48 \pm 0.04$.
This fitted curve is shown in Fig.~\ref{fig:lfsum}, and the
turnover particularly is very
close to our original expectation.  As discussed in Papers I and III and
previous papers on the GCLF \citep[e.g.][]{sec93,han87},
$V^{to}$ and $\sigma$ are correlated parameters and tend to be 
overestimated in cases where the data themselves do not reach
clearly past the turnover, since there are no $V > V^{to}$ observed bins 
to rule out solutions fainter than the true turnover level.
The combined GCLF is based effectively on a residual sample
after background subtraction of more than 12,000 GCs.  

For comparison, in Paper I we derived $V^{to} = 27.88 \pm 0.10$
for NGC 4874 
with a fit {\sl constrained} to $\sigma_V \equiv 1.4$ from an effective
sample of $\simeq 2200$ GCs; while in Paper III we obtained
$V^{to} = 27.8 \pm 0.2, \sigma_V = 1.5 \pm 0.1$ for IC 4051 from an effective
sample of $\simeq 2000$ GCs.  In both cases, the background was
a ``local'' one defined from the outer regions of each field itself,
and thus for NGC 4874 the GC population was somewhat oversubtracted.
The solution for the turnover level and dispersion in the present paper
is based on a GC sample five times larger than in either of these
previous studies, as well as an improved background definition. 
Both of these factors mean that
we can constrain the solution much better than in our previous work.
It is worth noting as well that our combined Coma sample of 
$\sim 10^4$ GCs brighter than the GCLF turnover is more than twice as large as
the numbers measured in all the Virgo galaxies combined \citep{pen06,jor07}.

The Gaussian is by far the most frequently used shorthand description
of the GCLF, but has no physical basis.
Other, more physically based, models have been proposed for the GCLF
that combine aspects of numerical convenience with some foundation
in a plausible initial mass function for GCs and its subsequent
dynamical evolution.  Descriptions of the LF in its alternate
form as number per unit luminosity ($dN/dL$, the {\sl luminosity
distribution function} or LDF) connect much more directly with
the mass distributions of young star clusters, and 
began to be applied to globular cluster populations in
several galaxies some years ago \citep[e.g.][]{hp94,mcl94}.  
Schechter-like functions have been applied to the high-mass
end of the globular cluster distribution by, e.g.,
\citet{bur00}, and to the observed distributions for young
massive clusters by \citet{gie06}, \citet{whi99}, and 
\citet{whi02} among others.  A similar form
has been developed further into an
evolved Schechter function formulation \citep{zha01,jor06,jor07} which accounts
roughly for the long-term dynamical evolution of GCs and
the preferential loss of low-mass clusters over time.
\citet{jor07} have tested this model extensively against the database
from the Virgo cluster survey over a wide range of host galaxy
luminosities.  Strictly on numerical grounds, this
function is better able to match the asymmetry of
the complete GCLF:  in the number-vs.-magnitude plot,
the faint end past the turnover is observed to have a broader
wing than the bright half.  However, for our purposes 
of defining the turnover point, this feature
of the model is irrelevant, since the Coma data fall far short of
the luminosity limit at which the differences between this function
and the simpler Gaussian become important.  
In Fig.~\ref{fig:lfsum}, we show our best-fit evolved Schechter
function in the form expressed by \citet{jor07},
\begin{equation}
N(V)  =  const { 10^{-0.4(V-V_c)} \over 
{[10^{-0.4(V-V_c)} + 10^{-0.4(\delta - V_c)}]^2} }
{\rm exp} [-10^{-0.4(V-V_c)}]  . 
\end{equation}
Our numerical solution gives $(\delta - V_c) \simeq 3.20$
and $V_c \simeq 24.40$, both uncertain to $\pm 0.05$ mag.
As is evident from Fig.~\ref{fig:lfsum}, there is no
distinguishable difference between the best-fit Gaussian
and Schechter-like functions {\sl for the luminosity range
around the turnover point and brighter}.  This bright-half
regime is what our data cover.  It is only on the unobserved
fainter half of the GCLF that the differences between the two
analytical approximations show up strongly.  
For the Schechter function fit, the magnitude $V_c$ above which the bright-end
exponential cutoff begins to dominate the shape of 
the LF corresponds roughly to a 
cluster mass $M_c \simeq 3 \times 10^6 M_{\odot}$ for $(M/L)_V
\simeq 2$ \citep{mcl00,rej07}.

The cutoff mass that we find is noticeably higher, and the overall LF
broader, than in any of the Virgo galaxies.  However,
\citet{jor07} already show from the Virgo sample
that $M_c$ increases systematically with
host galaxy luminosity.  Since our Coma GCS sample is
dominated by the supergiants NGC 4874 and NGC 4889, which are
more luminous than anything in Virgo,
an extrapolation upward from their correlation
of $M_c$ vs. $M_B^t$ puts the Coma result quite close to
what that trend would predict.

For our fiducial Coma distance $(m-M)_0 = 34.97 \pm 0.13$, 
then the observed turnover $V^{to} = 27.71 \pm 0.07$ converts
to $M_V^{to} = -7.32 \pm 0.13$, subtracting $A_V = 0.03$ and
a $V-$band K-correction of 0.03 (see Paper I and Frei \& Gunn 1994).  This turnover luminosity
agrees extremely well with the mean from many other giant
ellipticals in Virgo, Fornax, and other nearby groups of
$M_V^{to} = -7.3 \pm 0.1$ \citep[e.g.][and Paper I]{har01}.
Phrased differently, if we had chosen here to use $M_V^{to}$
for its classic purpose as a standard candle \citep{jac92,har01}
and to derive the distance to Coma, as we did in Paper I,
we would obtain $(m-M)_V = 35.01 \pm 0.15$.  
The corresponding Hubble ratio would be $h = 0.73 \pm 0.07$.
Most of the uncertainty in this case is actually from the local calibration
of the turnover luminosity.  

Perhaps the most relevant individual comparison is with M87 alone,
the central Virgo giant, whose GCS has been extensively 
surveyed to very deep levels in $(V,I)$. (A larger composite sample
is available from the Virgo cluster survey of \citet{jor07},
but these are in the $z'$ band.) \citet{wat06} present a
GCLF from WFPC2 data in the same $V$ and $I$ bands as we use
here, which extends well past the turnover level to high
completeness.  By subdividing their data into radial bins,
they show that the LF is also independent (to within the
statistical uncertainties) of location, so we use their
full data sample read from their Figure 4. We normalize
the M87 data to the same total GC population brighter than the
turnover level, which for M87 they find to lie at 
$V^{to}(M87) = 23.60 $ with $\sigma_V = 1.42$ for a Gaussian
fit; since we measure $\simeq 9000$ GCs to that level in
the combined Coma 4-galaxy sample, and they find $\simeq 519$
GCs to the turnover in M87, we multiply their bin totals
by 17.5.  For an adopted Virgo distance of 16 Mpc (Paper I)
and $A_V = 0.03$, this gives a relative distance modulus
$\Delta m = 4.04$ mag (again adding the 0.03-mag
K-correction at Coma).  Thus displaced to the Coma distance,
the M87 GCLF turnover would appear at $V^{to} \simeq 27.64$,
in good agreement with our data.  The normalized M87 distribution
is also shown in Fig.~\ref{fig:lfsum}; with 19 times fewer
GCs contained in the sample relative to the Coma data, 
it shows stochastic differences
from bin to bin that are distinctly larger, but there are
no clear systematic differences.  A standard Kolmogorov-Smirnov
two-sample test shows that there is a negligible probability that
they are intrinsically different for any magnitude range
brighter than the turnover region.

Lastly, our best-fitting GCLF dispersion of $1.48 \pm 0.04$ mag
fits into the general pattern that the dispersion increases
systematically with parent galaxy luminosity \citep{jor06,jor07}.
Our composite sample corresponds roughly to $M_B^t = -22.4$,
the average of NGC 4874 and 4889.
The mean relation $\sigma \simeq 1.12 - 0.093 (M_B^t + 20)$
derived by \citet{jor06} would then predict $\sigma = 1.34$
for our sample.  This is smaller than what we observe,
but the difference is within the galaxy-to-galaxy scatter around
the mean relation, which is at least $\pm 0.1$ mag (see their
Figure 1).

The two main conclusions from our LF analysis are that 
(1) the GCLF turnover luminosity is very similar to those in
other giant ellipticals; and (2) the bright half of the GCLF
may be significantly broader than in the Virgo members, as indicated by
either the Gaussian dispersion $\sigma$ or the 
Schechter-function parameter $(\delta-m_c)$.  We suggest that
this evidence favors a formation process in which the supergiants
did not assemble just as ``dry mergers'' of previously built
systems.  Because the GCLF is broader, the supergiants have
{\sl relatively} more high-mass GCs that seem more likely to
have been built in the gas-rich stages of hierarchical merging.
If GCs had simply been added later from gas-free satellite accretions,
it is difficult to see how the very highest-mass end of the
GCLF could have been populated to the extent that we see.

\section{Total Populations and Specific Frequency}

Knowing the results from the GCLF analysis and
the radial profile information, we can next estimate for each
galaxy the
total globular cluster population $N_t$  and specific frequency,
$S_N \equiv N_t \cdot 10^{0.4(M_V^t + 15)}$ \citep{har81}
where $M_V^t$ is the integrated magnitude of the galaxy.
To estimate $N_t$, we integrate under the GCS radial profiles
from Fig.~\ref{fig:lograd} and then multiply that total
by the fraction of the GCLF that is fainter than the 
observed limit.  Individual details follow.

\noindent {\sl NGC 4874:} Of the five galaxies in our study,
this is by far the hardest for which to calculate a global $S_N$.
To estimate the total GC population,
we use the radial range $4'' < r < 150''$
(almost the entire area of the WFPC2 field) and 
conservatively take the magnitude range $V < 26.5$ for
which the photometry is complete.  We further color-select
to include objects in the range $0.5 < (V-I) < 1.8$.
The background number density in this magnitude and color
range, as determined from the ``global'' background field,
is $\sigma_{bkgd} = 0.011$ arcsec$^{-2}$.
Over $4'' < r < 150''$, integration of the GC density profile gives
$3875 \pm 468$ clusters after background
subtraction.  Given that the innermost profile
is nearly flat (cf. Fig.~5), we use the value of $\sigma_{GC}$ at $4''$
and add 15 more GCs for the region inside that,
giving $N \simeq 3890 \pm 470$ GCs within $r = 150''$ of galaxy
center.  Next, we use the standard GCLF Gaussian shape determined
in the previous section to extrapolate the total over
all magnitudes.  The cutoff $V_{lim} = 26.5$ is $0.82 \sigma$ short
of the GCLF turnover at $V = 27.71$, which means that we multiply
our observed GC total by 4.81 to obtain the GC population over
all magnitudes (we note here that the specific frequency $S_N$
is {\sl defined} assuming a Gaussian GCLF:  that is, we find the number
of observed GCs to the turnover point, double that, and use the
resulting total to calculate $S_N$).  
This estimate yields $N_t = 18700 \pm 2260$.  
With $M_V^t = -23.43$, we obtain $S_N = 7.9 \pm 1.0$.
This value supersedes our original (lower) value in Paper II.
The increase results from a combination of a lower adopted background level
and an 0.15-mag correction to the $V$ magnitude zeropoint for
this galaxy (see Section 6 below).  Clearly, the best
way to refine the measurement of $N_t$ further 
is to obtain wider-field data and a better global background.

This estimate must, however, be only a lower limit to the true
GC population because it does not account for the numbers
of clusters that must be present
outside $r = 150''$ ($\sim 72$ kpc).  The obvious indication from
Fig.~\ref{fig:lograd} is that the GC density profile continues
well beyond.  We do not know how far to integrate outward,
but it should continue at least twice this far, to
$r \sim 300''$ (143 kpc); for comparison, the halfway point to
NGC 4889 is at $205''$ (100 kpc).
The $r^{1/4}$ profile shown in Fig.~\ref{fig:lograd} for the
outer region $r \gtrsim 20''$ is
roughly given by log $\sigma_{cl} = -0.4912 r^{1/4} + 0.2377$
where $r$ is in arcseconds and $\sigma_{cl}$ is in number of GCs
per arcsec$^2$.  Simply extrapolating this outward to $r = 300''$
as it is (which we have no obvious reason {\sl not} to do) would increase
the total GC population by a factor of $\simeq 2.2$; that is, there
are more GCs outside $r=150''$ than inside {\sl if} the same
profile holds true.  We would then have to increase our estimate
of the total GCS to $N_t \sim 41,000$ and the specific frequency
to $S_N \sim 17$.  Specific frequency values this high are certainly
within the range of measurements for other cD-type galaxies
\citep[see, for example,][]{weh08,tam06}.  

It is probably safe to say that NGC 4874 contains
more than 30,000 GCs, making it the most populous 
globular cluster system that we know of anywhere in the
local universe.  A more quantitative estimate
through wider-field imaging would be an extremely interesting
extension of the current study, and would also permit exploration of
the GC population within the Coma core region at large.

\noindent {\sl NGC 4889:}  For the second of the two supergiants, the
GC radial density curve is steeper.  For this reason, it approaches much more closely
to the background (Fig.~\ref{fig:lograd}) at the edges of the 
WFPC2 field, and so the the
estimate of $N_t$ can converge reasonably.  Using $V < 27$
and the global background, we obtain $\sigma_b = 0.040$ and 
a total of $3200 \pm 400$
clusters for radii $4'' < r < 150''$, and add another $\simeq 50$
for $r < 4''$.  Extrapolating over all magnitudes gives
$N_t = 11000 \pm 1340$. The 
outward extrapolation to account for clusters beyond the
observed field is a much smaller correction than for NGC 4874:
we find with $r(max)$ the same as used for NGC 4874 above
that it is necessary to increase $N_t$ by at
most 10-15\%, i.e. $N_t \simeq 12000 \pm 2000$.
With $M_V^t = -23.54$ we have $S_N = 4.7 \pm 0.6$.  
This result is squarely in the normal
range established by the giants in Virgo and Fornax; 
its huge total GC population simply goes along with its
very high luminosity.  

\noindent {\sl IC 4051:}  This GC-rich system fortunately has
an even steeper radial profile, so the limits on $N_t$ are even better.
With the same magnitude limit and background as above, we
obtain $2000 \pm 160$ for $4'' < r < 115''$, and we need
to add only about 10 more for the innermost region.
Integrating over all magnitudes gives $N_t = 6700 \pm 530$.
The radial profile is so close to background at this radius
that we do not add any further corrections; thus, with 
$M_V^t = -21.85$ we have $S_N = 12.1 \pm 1.0$, quite similar
to our result in Paper III.  Again we see the importance of
normalizing $N_t$ to the galaxy luminosity.  By sheer numbers,
the two central supergiants may contain most of the GCs in the
entire Coma cluster, but IC 4051 appears to have been at least
as efficient {\sl per unit luminosity}
as NGC 4874 and various cD galaxies elsewhere in forming them.
  
\noindent {\sl NGC 4881:} For $V < 26.0$ there are $\simeq 25$ GCs
after background subtraction and within radii $3'' < r < 30''$.  
To these we add about 25\% more as an estimate of the number of GCs lying in
the innermost and
outermost regions ($r < 3''$, $r > 30''$) not covered by our data.
Then, to correct the total for the GCs fainter than our adopted
limit of $V=26$, we note that this limit is 1.73 mag or 1.23 standard deviations
brighter than the mean GCLF turnover (previous section). This range includes
11\% of the true total cluster population.  We therefore scale up the observed
number by a factor of 9.  
With $M_V^t = -21.46$ (Table 1) we then obtain $S_N = 0.77 \pm 0.21$,
in very good agreement with the earlier estimate of $S_N \sim 1.0$ by \citet{bau95}.
This strikingly low$-S_N$ value is at the bottom end of the 
observed range for any type of E galaxy, regardless of luminosity 
or environment.  Our value may in fact even be a slight overestimate
if the galaxy actually has no clusters beyond the $30''$ (15 kpc) radius.

\noindent {\sl NGC 4926:} 
Again for the same limits as used above,
we calculate for NGC 4926 that for $4'' < r < 100''$
there should be $(570 \pm 110)$ clusters, and add in
$\sim 30$ more clusters for the innermost $4''$.
Extrapolating the radial profile outward gives at most 140 more
GCs, thus $N_t = 1300 \pm 300$ over all magnitudes as well.
With $M_V^t = -22.03$ we then have 
$S_N = 2.0 \pm 0.5$.  This total, on the low-to-normal side for E galaxies
\citep{har01}, is 6 times smaller than for IC 4051 but three times
larger than in NGC 4881.

Our final $N_t$ and $S_N$ estimates are listed in Table \ref{tab:targets}.
A completely independent set of specific frequency measurements
for the Coma galaxies has been made by \citet{mar02} through 
surface brightness fluctuations.  We have
three galaxies in common:  for NGC 4874, 4889, and IC 4051 they
derive $S_N = (9.0 \pm 2.2), (4.0 \pm 1.2)$, and $(12.7 \pm 3.2)$.
These numbers are in good agreement with our findings, to well
within their combined internal uncertainties.  It is also worth
noting that when the \citet{bla95} results
are adjusted for a 10\% larger Coma distance to bring them
in line with the current distance, their total populations would
become $\simeq 20700$ in NGC 4874 within $175''$ and 
$\simeq 15800$ in NGC 4889.  These are both consistent with
our estimates for the totals within $150''$.

\section{Color and Metallicity Distributions}

Lastly, we discuss the magnitude/color distributions a bit
further for the GCs in all five Coma galaxies.  Specifically,
we investigate whether or not the mean colors and color distribution 
fall into the now well-established pattern of bimodality
\citep[e.g.][among many others]{zep93,lar01,pen06,har06,kun07,str07}.
Although the $(V-I)$ index is not very sensitive to metallicity
\citep[see][]{geb99,kun01,lar01},
such data for many galaxies show a virtually universal pattern 
whereby the metal-poor mode is near $\langle V-I \rangle_0 \simeq 0.95$
and the metal-richer mode is near $\langle V-I \rangle_0 \simeq 1.15$,
with a weak trend for both modes to become redder with galaxy
luminosity \citep{lar01,str04,bro06} more or less as $Z \sim L^{0.15}$.
We would thus expect the GC modes for the Coma supergiants
particularly to be among the reddest known.

In Paper II we claimed that the mean colors for the GCs in NGC 4874
were bluer than normal for a giant elliptical, suggesting that most
of its cluster population was fairly metal-poor.  However, 
we discovered during the data reductions for the present paper that
the color scale of our photometry in Paper II   
was wrong by $\sim 0.15$ magnitude, resulting from the accidental use of
an incorrect color coefficient in the transformation from $F606W$ to $V$.
In the present paper, our NGC 4874 data (and indeed all the other 
galaxies) result from completely independent new reductions
starting from the raw images, and have all used the correct 
transformations.  As will be seen below, the true color distribution
for the NGC 4874 clusters indicates a normal combination of 
metal-poor and metal-rich clusters.

\begin{figure}
\plotone{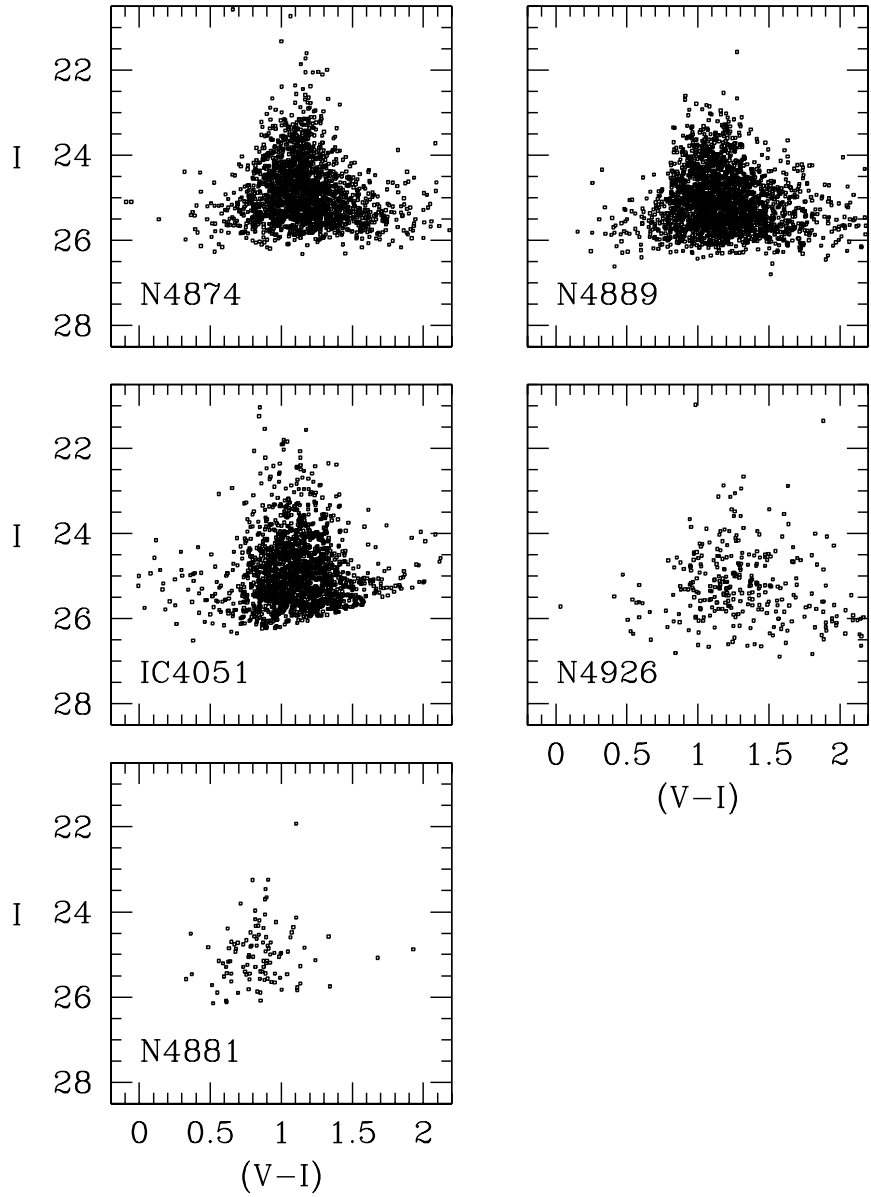}
\caption{Color-magnitude plots for the measured starlike objects
around the five galaxies.  All such objects on each WFPC2 field
are shown.
\label{fig:cmd5}}
\end{figure}

\begin{figure}
\plotone{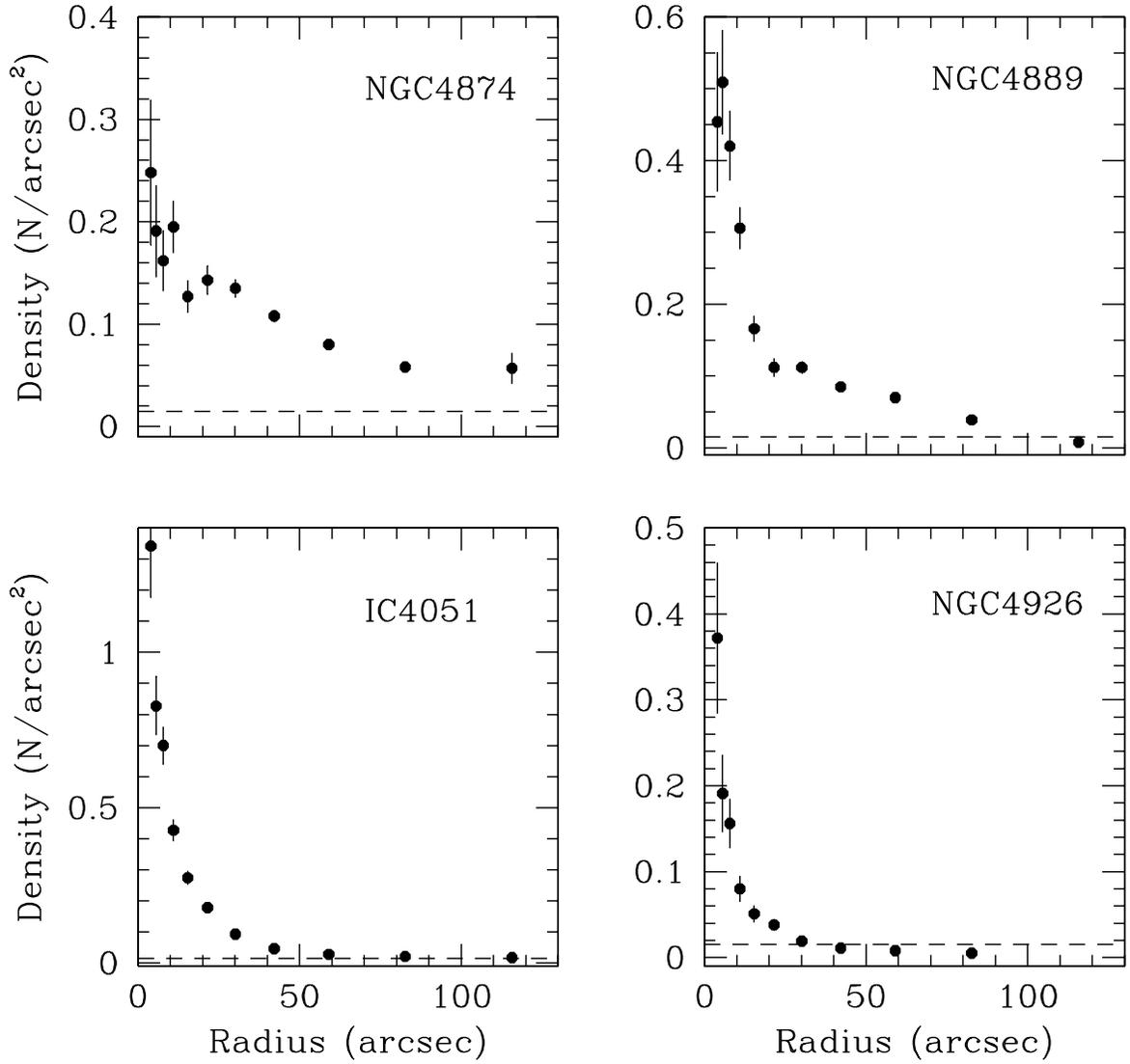}
\caption{Radial distributions for the measured starlike objects around
the four galaxies labelled.  Number of measured objects per arcsec$^{-2}$
is plotted against radius from galaxy center; the linear scale is
50 kpc per $100''$.  Here the sample is restricted to objects
in the range $22.0 < I < 25.5$ and $0.5 < (V-I) < 1.8$, which includes
the brighter part of the globular cluster systems.  The approximate
``background'' level of field contamination is 
shown by the dashed lines.  Notice the different
vertical scales on each graph; for example, although NGC 4874
has by far the most spatially extended GCS of all these galaxies,
continuing well beyond the border of our WFPC2 field,
it does not have the GCS with the highest central density.  
\label{fig:rad4}}
\end{figure}

\begin{figure}
\plotone{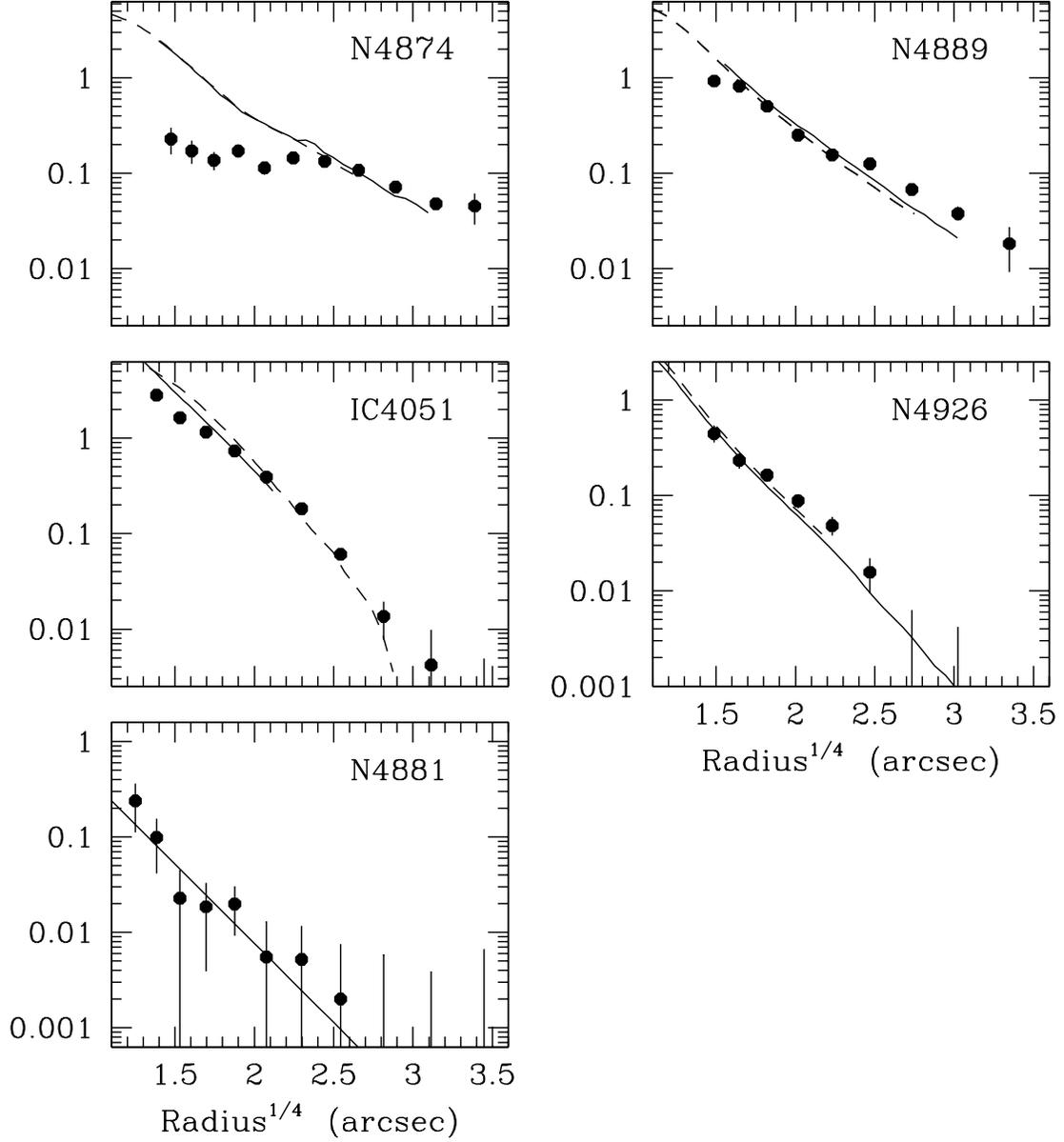}
\caption{Radial distribution of the globular cluster systems
around our five galaxies.  The ordinate of each graph is the
number density, $\sigma_{cl}$ of globular clusters per
arcsec$^2$.  The $r^{1/4}$ values are for radius in arcseconds.
In each graph the lines give the radial falloff of the surface
light intensity of each galaxy; see text for sources.
\label{fig:lograd}}
\end{figure}

\begin{figure}
\plotone{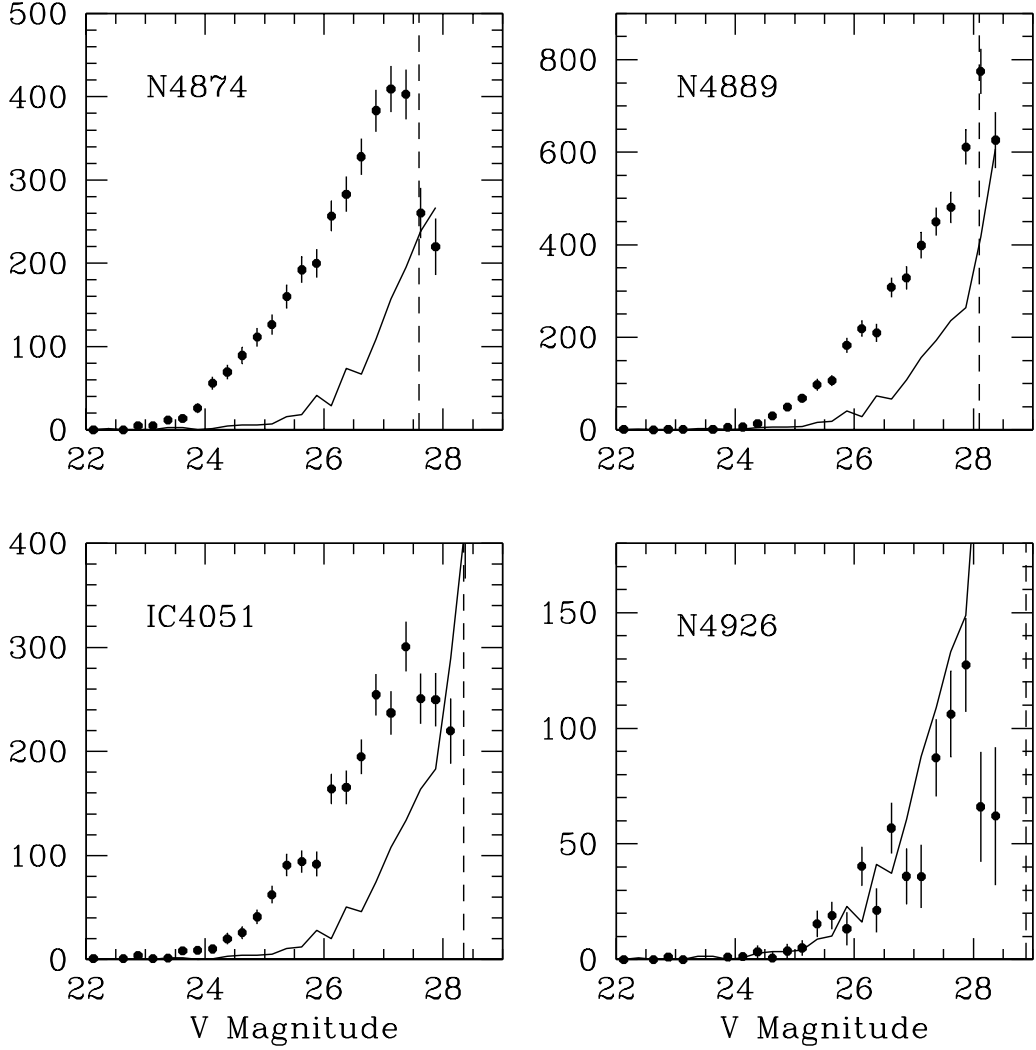}
\caption{Luminosity functions for the globular clusters shown
individually for the four Coma galaxies with our deepest photometry.
The LFs are {\sl after} background subtraction and completeness-corrected.
For comparison, the adopted background LFs are shown as the
broken line.
Number of clusters per 0.25-mag bin is plotted against $V$.
In each panel, the vertical dashed line indicates the 50\% 
completeness level of the photometry in the WF2,3,4 frames.
\label{fig:lf}}
\end{figure}

\begin{figure}
\plotone{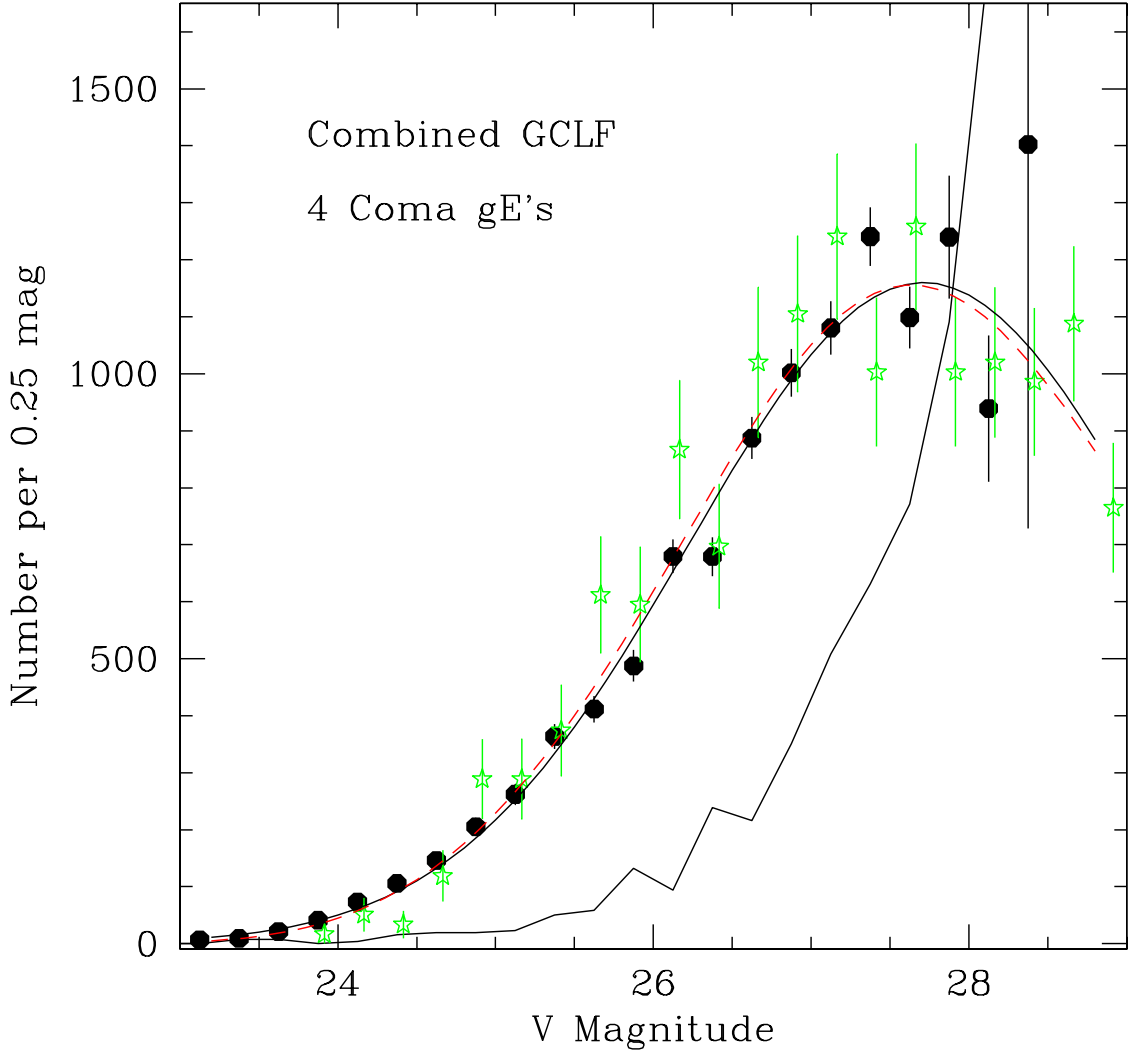}
\caption{Composite luminosity function for the globular clusters
in four Coma ellipticals, shown in solid dots with error bars.
The composite LF is completeness-corrected and background-subtracted
according to the prescription described in the text.
The broken line at bottom is the background LF by itself, for comparison.
The GCLF for the Virgo giant M87 is shown for comparison in
open starred symbols, shifted by $\Delta V = 4.04$ mag and normalized
to the same total GC population brighter than the turnover point (see text).
The best-fitting Gaussian function, with turnover level $V^{to} = 27.71$ 
and dispersion $\sigma_V = 1.48$, is shown as the solid line. 
The dashed line gives the best-fit evolved Schechter function model,
which has parameters $(\delta-V_c) = 3.2$ and a ``cutoff'' magnitude $V_c = 24.4$.
\label{fig:lfsum}}
\end{figure}

\begin{figure}
\plotone{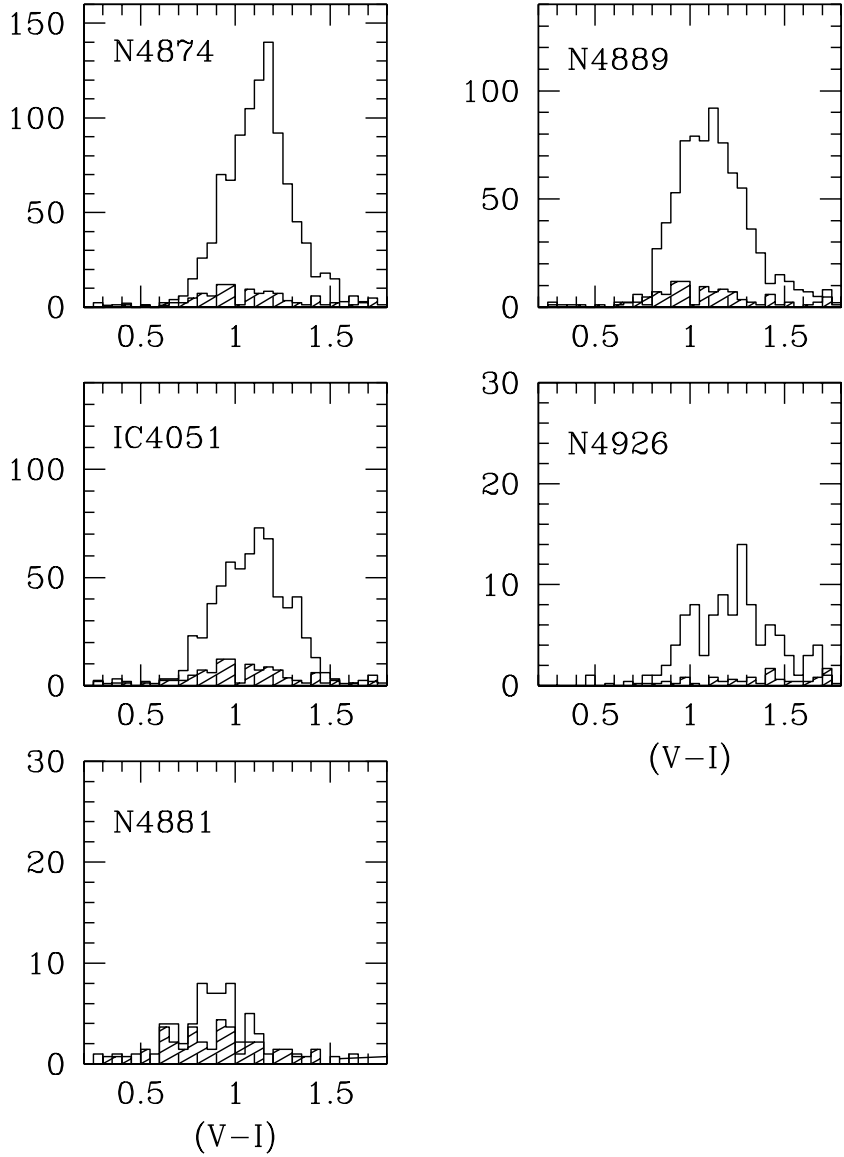}
\caption{Color histograms in $(V-I)$ for the starlike sources around
the five galaxies.  In each case the ``background'' population at large
radius is shown as the hatched histogram, while the population of objects
closer to the galaxy is the open histogram (not background-subtracted).  
Both histograms are normalized
to the same area on the sky.  For the first three objects 
(NGC 4874, NGC 4889, IC 4051) the background color distribution is the
mean of the far-field distributions from IC 4051 and NGC 4926, as described
in the text.  For the last two objects (NGC 4926, NGC 4881) the background is
a locally defined one (see text).
\label{fig:vihisto}}
\end{figure}

The color-magnitude data in Fig.~\ref{fig:cmd5} show that the
raw $(V-I)$ colors fall in the normal range for GCs
in giant E galaxies, except for NGC 4881
in which the GCs appear to be almost entirely blue.  In the other four,
the standard red and blue modes (if present) are substantially blurred 
and overlapped by photometric measurement scatter.  
We restrict our discussion of the color histograms to the relatively
bright range $I < 25.5$, well above the GCLF turnover point.  
These are shown in Figure \ref{fig:vihisto}.

\begin{figure}
\plotone{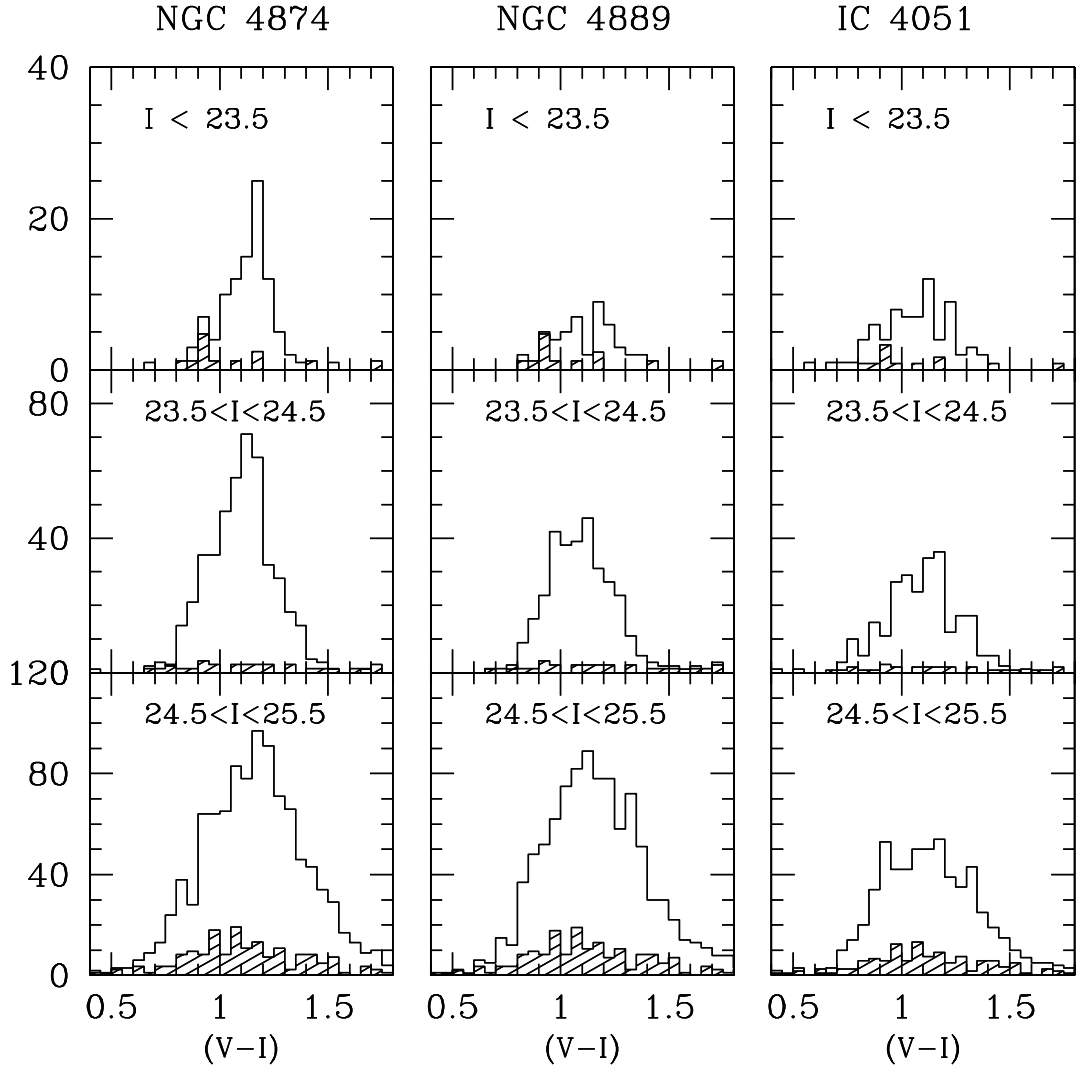}
\caption{Color histograms for the globular clusters around the three most
populous systems in our study, now subdivided by magnitude range.
The $(V-I)$ distribution of stars in the global background field is
shown in the filled histograms.  These should be subtracted from the
open histograms to give the residual GC population distributions.
\label{fig:vi3panel}}
\end{figure}

\begin{figure}
\plotone{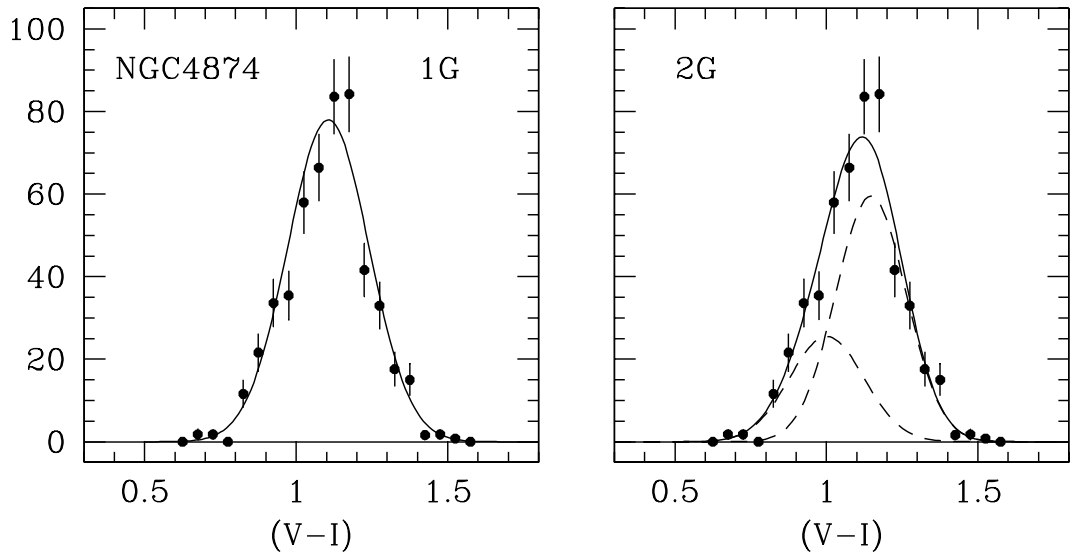}
\caption{Color histogram for the bright globular clusters around NGC 4874
($I < 24.5$) for which field contamination and photometric errors are
minimal.  The left panel shows a best-fit single Gaussian function to
the color histogram, while the right panel shows a {\sl possible}
two-Gaussian solution to the same data, with the parameters as listed
in Table \ref{tab:cfit}.
\label{fig:cfit4874}}
\end{figure}

\begin{figure}
\plotone{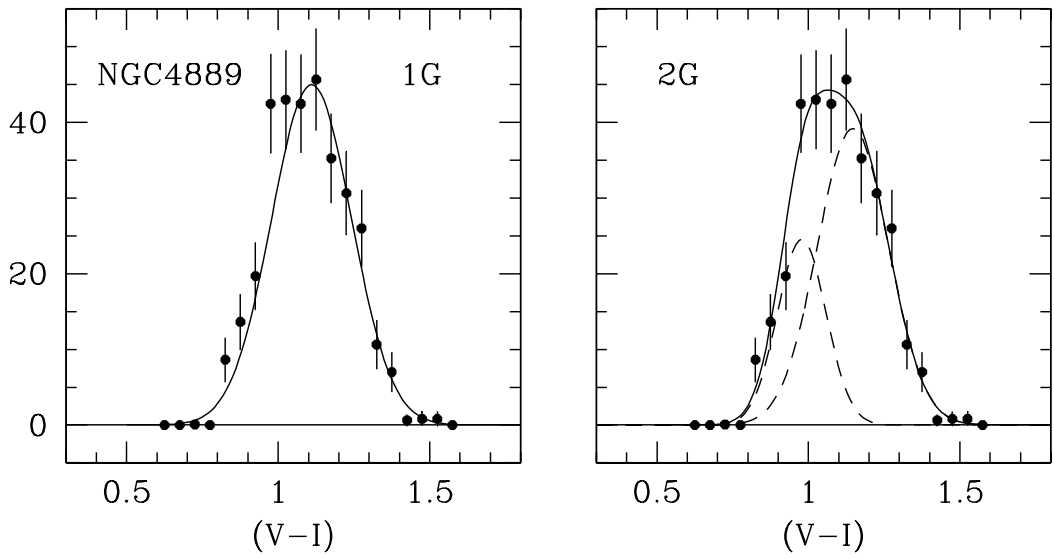}
\caption{Same as the previous figure, for the bright clusters around
NGC 4889.  In this case the two-Gaussian model provides a significantly
better fit.
\label{fig:cfit4889}}
\end{figure}

\begin{figure}
\plotone{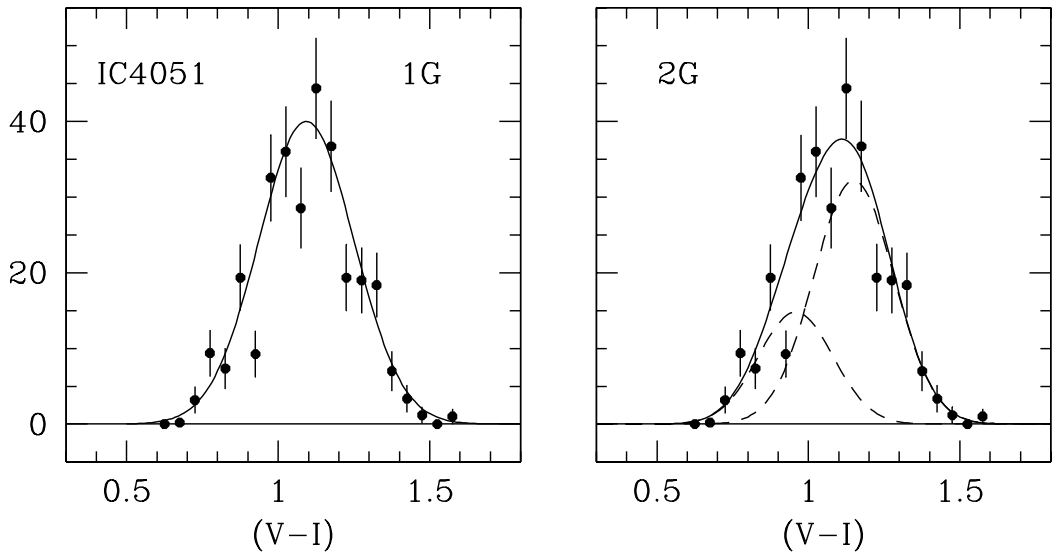}
\caption{Same as the previous figure, for the bright clusters around
IC 4051.  
\label{fig:cfit4051}}
\end{figure}

For the three most populous GCSs, we can go a bit further and
look at the same histograms broken down by magnitude range, as
shown in Figure \ref{fig:vi3panel}.  The numbers of stars in the
global background field in the same ranges, normalized to
the same area as each galaxy field, are shown in the shaded
histograms.  In order to minimize the damaging effects of
both field contamination and color spread from photometric
measurement scatter, we choose to use only the objects in
the very bright range $I < 24.5$ (corresponding roughly to
GC luminosities $M_V \lesssim -9.5$) for further analysis.
After subtracting off the small amount of background,
we obtain the histograms in
Figures \ref{fig:cfit4874}, \ref{fig:cfit4889}, and \ref{fig:cfit4051}, 
which show our best estimates for the
residual color distributions of the GCs alone.

Do these distributions show the now-standard bimodal form?
Even with our most rigorous subselection of data, the answer
is ambiguous, and in the end we can gain only some hints.
We use the statistical package RMIX \citep{mac07,weh08} to
perform multimodal fitting to each histogram, and specifically
to gauge whether or not two modes might be present.
Although RMIX can use a variety of fitting functions, we stick
with the Gaussian model for simplicity and ease of comparison
with the previous literature.  (See Wehner et al.~2008 for examples of
its use in the same context of GCS color distributions.)
In the left panels of
Figs.~\ref{fig:cfit4874}, \ref{fig:cfit4889}, and \ref{fig:cfit4051}, 
we show the best-fit \emph{unimodal} Gaussian curve for each of
the three galaxies, with means and
standard deviations listed in Table \ref{tab:cfit}. In all cases,
these provide at least roughly adequate descriptions of the run of
the data.  If we next ask RMIX to find two modes with complete freedom
to solve for the means, standard deviations, and proportions, the code tends to
converge to solutions not far from the single-Gaussian solution, i.e.
ones in which one of the modes takes up $\lesssim$ 10\% of the population.
In short, without external arguments to the contrary, we have no
compelling evidence to claim the presence of bimodal sequences.

Instead, we ask a slightly more restricted question.
{\sl If the red and blue modes are actually present} and simply being
obscured by photometric scatter, we would like to know
most importantly their mean colors $\mu_1, \mu_2$
and the proportions $p_1, p_2$.  To estimate these, we can
press the data a little further by performing
constrained bimodal fits where we assume fixed input values
for various combinations of the parameters, such as 
the dispersions $\sigma_1, \sigma_2$.
We have experimented widely with various combinations
of these constraints, which are straightforward to do within RMIX.
Samples of these are shown in the right panels of
Figs.~\ref{fig:cfit4874}, \ref{fig:cfit4889}, and \ref{fig:cfit4051}.
These are not intended to be our ``best'' choices 
because ones producing equally good fits can be obtained with other pairs
of parameters (although we
justify our choice of the 30/70 proportions a bit further
below). They are, however, illustrative of
the quality of fit that the constrained bimodal solutions provide.
Any other solutions in which either $p_1 \sim p_2$
or $\sigma_1 \sim \sigma_2$
yield quite similar means $\mu_1, \mu_2$ to the
ones shown here; that is, the solutions for the mean colors
tend to be robust against plausible changes in either the
proportions or dispersions.  The sample two-Gaussian solutions are 
listed in Table \ref{tab:cfit} for these three galaxies, in the 
pair of lines just below the single-Gaussian data.  In each case the
double-Gaussian solution is slightly but not strongly preferred
over the single-Gaussian.

Perhaps the most interesting single conclusion from these admittedly
rough tests is that the deduced mean
color of the blue mode is virtually identical at 
$\langle \mu_1 \rangle = 0.98 \pm 0.02$
in all three systems.  The same is true for the red mode, where
$\langle \mu_2 \rangle \simeq 1.15 \pm 0.02$ in all three systems.
Subtracting off the foreground reddening of just 0.01 mag,
and a K-correction $K_V-K_I \simeq 0.02$ \citep[e.g.][]{fre94},
we have $\mu_0$(blue) = 0.95, $\mu_0$(red) = 1.12, which are
quite close to the previously known values for the brightest giant ellipticals
elsewhere \citep{lar01,str04,bro06} and 
agree with these to well within the uncertainties
of the data.  These two arguments reinforce our
tentative conclusion that the normal two modes may be present
in these Coma members, and lie at least approximately at 
the expected metallicity values.

\begin{figure}
\plotone{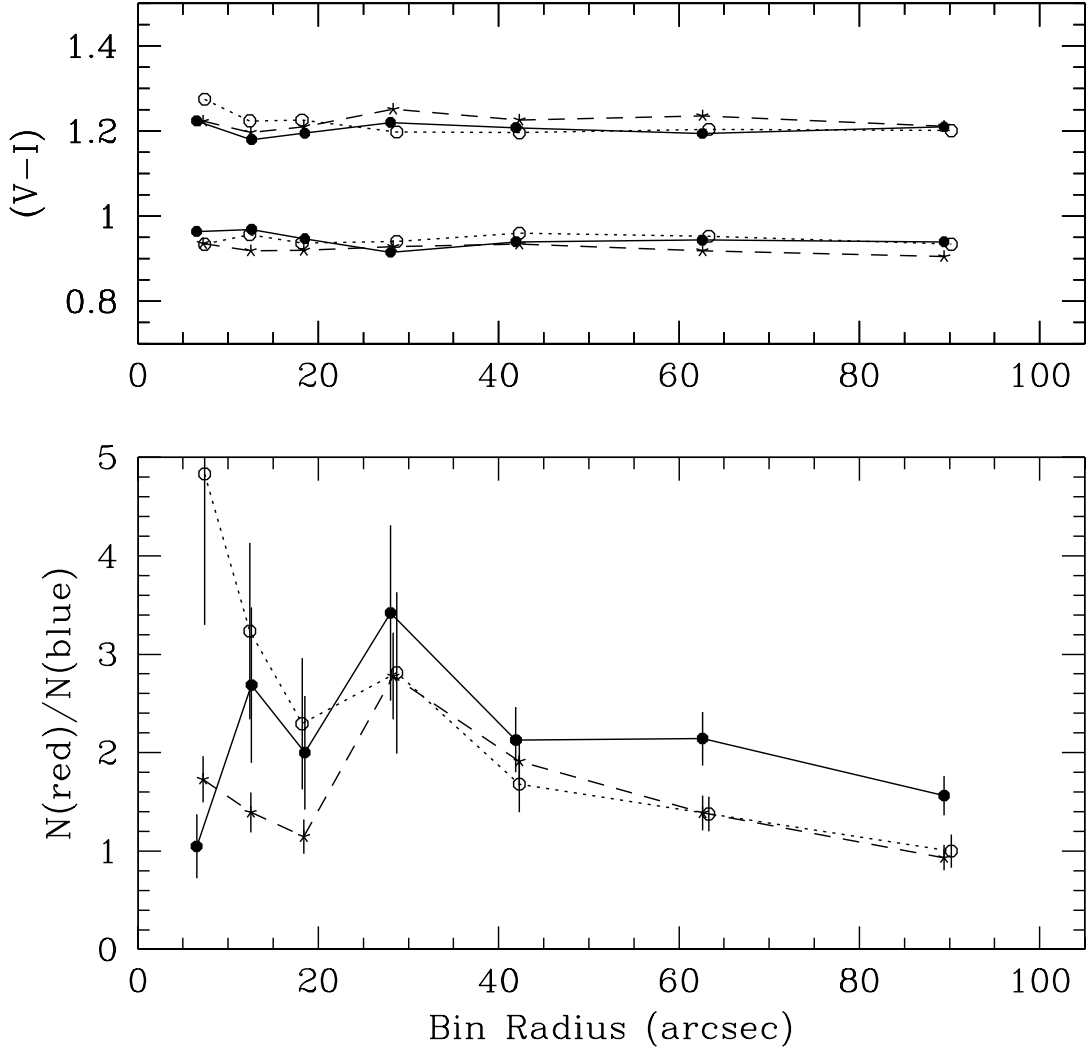}
\caption{{\sl Upper panel:}  Mean color of the blue GCs (lower set of three
lines) and the red GCs (upper set of three lines) as a function of
radius from galaxy center.  Data for NGC 4874 are the solid dots and
lines; NGC 4889 the squares and dotted lines; and IC 4051 the stars and
dashed lines. No color gradient within either the blue or red populations
is detectable.  
{\sl Lower panel:}  Ratio of number of red GCs to number of blue GCs,
plotted as a function of radius.  Solid dots are for NGC 4874, open
circles for NGC 4889, and stars for IC 4051.  A slight population
gradient exists for each galaxy.
\label{fig:colorgrad}}
\end{figure}

One further test of these three color distributions, now by radius,
is shown in Figure \ref{fig:colorgrad}.  If two subpopulations of
GCs are present in blue and red sequences, then evidence from
other galaxies indicates that we may expect to
find an overall population gradient with galactocentric distance
in the sense that the redder, more metal-rich GCs are more
centrally concentrated.  As a rough dividing line, we simply take
GCs brighter than $I=25$ 
with $(V-I) < 1.05$ and put them into the ``blue'' population,
and ones with $I < 25, (V-I) > 1.05$ into the ``red'' population.  
Fig.~\ref{fig:colorgrad} shows the result of two numerical experiments.
In the first panel we show the mean color of the red and blue
subsamples in radial bins.  We conclude that no net change in
the mean color of either group is present.  (As noted above from
the bimodal Gaussian fits, the mean colors of each group are also
closely similar in all three galaxies.)

In the second panel we show the population {\sl ratio} $N(red)/N(blue)$
in the same radial bins.  A slight but noticeable trend appears 
in all three galaxies in the expected sense, for the more metal-rich
clusters to dominate more in the inner regions.  This trend is
most noticeable for NGC 4889 (dotted line) but for $r > 30''$ it is
present almost equally in all three.  Overall the red
ones make up more than 2/3 of the total (justifying {\sl post facto}
our choice of the 30/70 proportions in the Gaussian fits described above).
However, at very large radius (40 kpc and more) the ratio begins
to approach $N(blue) \simeq N(red)$.  
We add this evidence to support our tentative
conclusion that a radial population gradient does exist in these
galaxies:  the mean metallicities of both the two modes are
roughly constant with radius, but their relative numbers change.
Very similar evidence has been presented for giant
ellipticals in Virgo, Fornax and other Abell clusters
\citep{gei96,dir03,rho04,har06}.

In all three of
NGC 4874, 4889, and IC 4051 we find that more than half
the clusters fall in the metal-rich mode {\sl at any radius
within our surveyed area.}  If these proportions 
-- which we emphasize are admittedly still internally uncertain -- are
confirmed by more precise data where the two modes can
be split more definitively, it is worth noting that 
they would disagree with 
the trend noted by \cite{pen08}.  For the Virgo galaxies,
Peng et al.~find essentially that the ratio $N(red)/N(blue)$ increases
steadily as we pass along the sequence 
from dwarfs to giants, reaching a
ratio near $\sim 0.8$ for moderately large ellipticals (see their
Figs. 8 and 9). But the ratio 
then starts to decrease again down to $\simeq 0.5$ or less for the biggest Virgo giants
(M87, M49 and a few others).  Our three Coma supergiants, for which 
we find tentative evidence that $N(red)/N(blue) \gtrsim 1$,
may provide new counterexamples to this trend, perhaps
suggesting that at the very densest and most massive protogalactic
environments, efficient and 
high-metallicity cluster formation was especially favored.
To put this result another way,
whereas \citet{pen08} found that the galaxy-to-galaxy variations
in specific frequency in Virgo are driven mostly by the numbers
of {\sl blue} clusters, the similarly high specific frequencies
in the Coma supergiants are produced mostly by their large numbers
of {\sl red} clusters.

One final note about the color distributions relates to their
behavior at the bright end.  \citet{weh08} find, for the Hydra cD
galaxy NGC 3311, that the red GC sequence extends distinctly further
upward than does the blue sequence.  The very brightest red GCs extend
well into the UCD-type luminosity regime, suggesting a connection between
the most massive GCs and UCDs.  In our Coma data, we see a hint of such
a feature in the color-magnitude diagrams for NGC 4874 and IC 4051
(Fig.~\ref{fig:cmd5}).  For $I < 23$ (corresponding to $M_I < -12$), much more
than half the GCs are on the red side at $(V-I) \simeq 1.2$.
More extensive areal coverage of the NGC 4874 system particularly should
reveal whether or not this effect is real, or simply an accident
of small-number statistics.  In addition, a larger sample of red, very luminous
clusters would allow an interesting test of the recent model by
\citet{bai08}, which predicts that a modest but real increase of cluster
metallicity with mass
should appear at this top end.  This mass/metallicity
relation (MMR) in their model is driven by self-enrichment during cluster
formation and should affect the highest-mass clusters along \emph{both} the
blue and red sequences, but is more noticeable along the blue
sequence \citep{har06,str06,mie06} because initial (pre-enrichment) metallicity
level is much lower.

\begin{figure}
\plotone{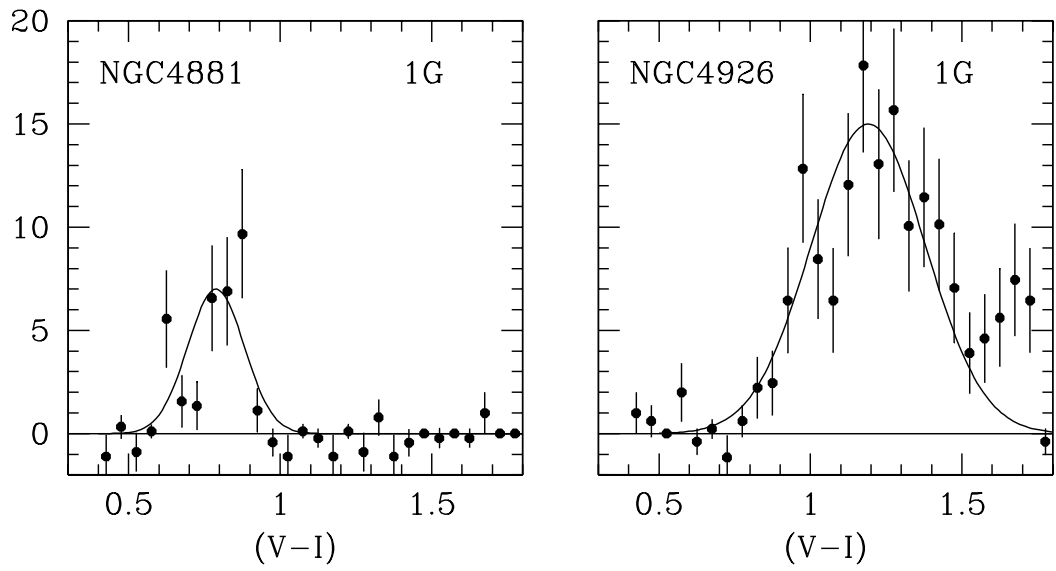}
\caption{Color histograms for the GCs around NGC 4881 and 4926,
as described in the text.  Single-Gaussian fits are shown for each.
\label{fig:cfit2}}
\end{figure}

For NGC 4881 and 4926, the GC populations are too small to
attempt more than single-mode descriptions.  In Figure \ref{fig:cfit2}
we show the results for these two low$-S_N$ systems, with numerical
solutions as listed in Table \ref{tab:cfit}.
Guided by the radial density plots in Fig.~\ref{fig:rad4},
we used local background subtraction to define the residual color
histograms for these two systems.
For NGC 4881, the local background was adopted to be the region
$r > 40''$ from galaxy center, while for NGC 4926 it was $r > 50''$.

The color distribution for
NGC 4881 is perhaps the most surprising one in our entire study,
falling farther away from
the normal pattern than any of the others in our sample.  The few GCs that it has
are blue, with a mean color $\mu = 0.79$  and rather narrow
histogram.  The correlation of blue-sequence
$(V-I)$ color with galaxy luminosity by \citet{str04} would predict
$\mu \simeq 0.94$, far redder than what we see; mean colors as low
as $\simeq 0.8$ are typically found only in dwarf ellipticals
or in the halos of spirals like the Milky Way.
Both the low $S_N$ of this galaxy and its very metal-poor GC
population are severe challenges to normal models for formation
of large ellipticals.  If it did form through any series of mergers,
these must have been virtually gas-free to avoid making any more
metal-richer clusters, but furthermore, the progenitor systems must
have been lacking such clusters to begin with.  But if the original
mergers of small progenitors were that gas-free, it would have been
even more difficult to form the metal-rich field stars that make up
its bulk population.  The mean color of the galaxy as a whole
of $(B-V))0 \simeq 1.00$ is in the normal range for metal-rich gE's
\citep[see also][]{tho87}.  This remains an intriguing system.

Finally, NGC 4926 presents almost the opposite case to puzzle over.
The color histogram is fairly broad and has a mean color $\mu = 1.19$
that is a bit redder than even the red sequences discussed above
for the supergiants.  
The small excess of
very red objects with $(V-I) > 1.5$ is mostly clustered in one
quadrant of the WFPC2 image and probably is due to a background
cluster of galaxies.
If there is a blue GC sequence (perhaps centered
at $(V-I) \simeq 0.95$; see Fig.~\ref{fig:cfit2}) it would resemble
the blue modes we found for the three major systems above, though the
statistics are too uncertain to allow firm claims.  
Rapid early mergers, at a time when a large amount of gas was
available for star formation, might be capable of producing
a GC abundance distribution like this, but it is less obvious
how the same series of events could have left behind a rather low
total population of clusters.

We will leave the discussion at this point.  Much improvement in
understanding the GC metallicity distributions in these galaxies
can be expected from ACS/WFC imaging in the near future \citep{car08}.

\section{Summary}

In this paper, the last of our series on the Coma galaxies, we
have presented our analysis of WFPC2 imaging of the globular
cluster systems around five giant ellipticals.  Our principal
results can be summarized as follows:

\begin{itemize}
\item{} The supergiant cD galaxy NGC 4874 holds
perhaps more than 30 thousand clusters;
its true radial extent is not yet known but may fill up a substantial
part of the Coma core.  It is the most populous globular
cluster system in any galaxy that we know of.  Nevertheless, the radial structure analysis
shows that its GCS belongs to the galaxy rather than the Coma
potential well as a whole (cf. Paper II).
\item{} The range of specific frequencies we see in just these
five galaxies is amazingly large, ranging from a low of $S_N < 1$
in NGC 4881 up to $S_N \simeq 12$ in both NGC 4874 and IC 4051. 
Their past histories of tidal truncation, gas-free
or gas-rich mergers, or GC formation efficiency may all have played
roles, but puzzles remain that simply do not have clear explanations
as yet.
\item{} Our $V-$band data for four galaxies (NGC 4874, 4889, 4926, 
IC 4051) reach deep enough to allow us to study the bright half
of the GC luminosity function thoroughly, and to define the
classic turnover point with some confidence.  We find 
$V^{to} = 27.71 \pm 0.07$, corresponding to $M_V^{to} = -7.3$
and quite similar to most normal E galaxies.  Our definition
of the GCLF is based almost $10^4$ GCs brighter than
the turnover point.  We find that both a simple Gaussian curve
and an ``evolved Schechter function'' fit these bright parts
of the GCLF equally well.  Though the turnover luminosity is
the same as in many other giant ellipticals, the GCLF shape
is noticeably broader, extending to quite high GC mass.
\item{} For the three biggest GC systems (NGC 4874, 4889, IC 4051),
analysis of the $(V-I)$ color distributions shows that all
three populations are dominated by red, metal-rich clusters.
Finer analysis is hampered by the random scatter in the 
photometry.
However, various tests including bimodal fitting to the color
histograms, and measurement of the population ratios (red vs.
blue GCs as a function of radius), show clear hints that 
the two normal color modes exist in these three systems, at mean colors
$\langle V-I \rangle_0$(blue) $\simeq 0.95$ and $\langle V-I \rangle_0$(red)
$\simeq 1.12$.  These values fall along the previously established
correlations of mean color with galaxy luminosity.
\item{} The three non-supergiant ellipticals in our study, NGC 4881, NGC 4926,
and IC 4051, present opposing challenges
to understanding their formation.  NGC 4881 has few GCs and these
have entirely blue colors like those in dwarf ellipticals; it
is completely lacking the metal-rich GCs that we conventionally
find in other big galaxies of all types.  NGC 4926 also has a
low $S_N$, and is far from the most luminous Coma member, but
is dominated by GCs that are at least as red as those in
the supergiants.  Finally, IC 4051 has a $S_N$ value as high as
many cD-type central giants (and more than an order of
magnitude larger than in NGC 4881), yet is structurally quite similar
in other ways to NGC 4881 and 4926.  

No single formation scenario
seems to be able to account for the huge range of characteristics
we see in all of these GCSs.
\end{itemize}

\acknowledgements

This research was supported through grants to WEH, DAH, and CJP from
the Natural Sciences and Engineering Research Council of Canada.
We are grateful to John Blakeslee for a careful and constructive
reading of the first version of the paper.

{}

\clearpage

\begin{deluxetable}{lccccr}
\tablewidth{0pt}
\tablecaption{Coma Target Galaxies }
\tablehead{
\colhead{Galaxy} & \colhead{$M_V^t$ } & \colhead{Velocity } & 
\colhead{Radius } & $N_{tot}(GC)$ & $S_N$ \\
\colhead{} & & \colhead{(km/s)} & \colhead{(arcmin)} \\
}
\startdata
 NGC 4874 & $-23.43$ & 7224 & 0 & $>18700 \pm 2260$ & $>7.9 \pm 1.0$ \\
 NGC 4881 & $-21.76$ & 6740 & 6.5 & $290 \pm 80$ & $0.8 \pm 0.2$ \\
 NGC 4889 & $-23.67$ & 6495 & 7.2 & $11000 \pm 1340$ & $4.7 \pm 0.6$ \\
 NGC 4926 & $-22.20$ & 7887 & 36.5 & $1300 \pm 300$ & $2.0 \pm 0.5$ \\
 IC 4051 & $-22.06$ & 8793 & 17.5 & $6700 \pm 530$ & $12.1 \pm 1.0$ \\
\enddata
\label{tab:targets}
\end{deluxetable}

\begin{deluxetable}{lcccc}
\tablewidth{0pt}
\tablecaption{Raw Image Data }
\tablehead{
\colhead{Galaxy} & \colhead{$V$ Filter} & \colhead{Exposure } & 
\colhead{$I$ Filter} & \colhead{Exposure}\\
}
\startdata
 NGC 4874 & $F606W$ & 20940 s & $F814W$ & 8720 s \\
 NGC 4881 & $F555W$ & 7200 s & $F814W$ & 7200 s \\
 NGC 4889 & $F606W$ & 23400 s & $F814W$ & 7800 s \\
 NGC 4926 & $F606W$ & 31200 s & $F814W$ & 10400 s \\
 IC 4051 & $F606W$ & 20500 s & $F814W$ & 5200 s \\
\enddata
\label{tab:data}
\end{deluxetable}

\begin{deluxetable}{lcccc}
\tablewidth{0pt}
\tablecaption{Completeness Function Parameters}
\tablehead{
\colhead{Galaxy/ccd} & \colhead{$\alpha_V$} & \colhead{$m_0(V)$ } & 
 \colhead{$\alpha_I$} & \colhead{$m_0(I)$ } \\
}
\startdata
 NGC 4874/PC1 & 3.7 & 27.60 & 4.0 & 25.20 \\
 NGC 4874/WF2,3,4 & 3.7 & 27.60 & 3.6 & 25.75 \\
 NGC 4881/PC1 & 4.3 & 26.15 & 4.6 & 25.30 \\
 NGC 4881/WF2,3,4 & 4.5 & 26.70 & 3.6 & 25.70 \\
 NGC 4889/PC1 & 1.5 & 28.00 & 2.3 & 25.75 \\
 NGC 4889/WF2,3,4 & 2.3 & 28.10 & 3.5 & 25.90 \\
 NGC 4926/PC1 & 3.3 & 28.15 & 4.6 & 25.80 \\
 NGC 4926/WF2,3,4 & 1.8 & 28.89 & 2.6 & 26.60 \\
 IC 4051/PC1 & 2.6 & 27.70 & 3.5 & 25.40 \\
 IC 4051/WF2,3,4 & 2.5 & 28.35 & 3.6 & 26.10 \\
\enddata
\label{tab:completeness}
\end{deluxetable}

\begin{deluxetable}{cccc}
\tablewidth{0pt}
\tablecaption{GCS Radial Profile Data }
\tablehead{
\colhead{$\langle r\rangle$} & \colhead{n} & \colhead{Area } & \colhead{$\sigma$} \\
\colhead{(arcsec)} & & \colhead{(arcsec$^2$)} & \colhead{(arcsec$^{-2}$)} \\
}
\startdata
NGC 4881 \\
   2.45 &  4 & 15.8 & $0.253 \pm 0.126$ \\
   3.67 &  4 & 35.2 & $0.114 \pm 0.057$ \\
   5.51 &  3 & 79.5  & $0.038 \pm 0.022$ \\
   8.27 &  6 & 179.1 & $0.034 \pm 0.014$ \\
  12.40 & 14 & 402.4 & $0.035 \pm 0.009$ \\
  18.60 & 15  &  733.1 & $0.021 \pm 0.005$ \\
  27.90 & 24  & 1185.6 & $0.020 \pm 0.004$ \\
  41.85 & 49  & 2888.2 & $0.017 \pm 0.002$ \\
  62.78 & 96 & 6212.8 & $ 0.016 \pm 0.002$ \\
  94.17 & 81 & 5925.4 & $ 0.014 \pm 0.002$ \\
 141.25 &  2 &  168.1 & $ 0.012 \pm 0.008$ \\
\\
NGC 4874 \\
   4.73 & 12 & 48.5 & $0.248 \pm 0.072$ \\
   6.63 & 18 & 94.4 & $0.191 \pm 0.045$ \\
   9.28 & 28& 185.6 & $0.151 \pm 0.029$ \\
  12.99 & 58  & 315.3 & $0.184 \pm 0.024$ \\
  18.18 & 54  & 497.0 & $0.109 \pm 0.015$ \\
  15.45 & 88  & 758.3  & $0.116 \pm 0.012$ \\
  35.64 & 192 & 1585.8 & $0.121 \pm 0.009$ \\
  49.89 & 305 & 3212.8 & $0.095 \pm 0.005$ \\
  69.89 & 383 & 5578.7 & $ 0.067 \pm 0.004$ \\
  97.79 & 262 & 4900.2 & $ 0.053 \pm 0.003$ \\
 131.74 & 15 &  264.4 & $ 0.057 \pm 0.015$ \\
\\
NGC 4889 \\
   4.90 & 61 & 63.2 & $0.966 \pm 0.124$ \\
   7.35 & 121& 141.1& $0.857 \pm 0.078$ \\
  11.02 & 173& 318.1 & $0.544 \pm 0.041$ \\
  16.53 & 168& 575.8 & $0.292 \pm 0.022$ \\
  24.80 & 145& 738.7 & $0.196 \pm 0.016$ \\
  37.20 & 348 & 2103.8 & $0.165 \pm 0.009$ \\
  55.80 & 492 & 4572.9 & $0.108 \pm 0.005$ \\
  83.70 & 583 & 7490.5 & $0.078 \pm 0.003$ \\
 125.56 & 59 & 1012.8 & $ 0.058 \pm 0.008$ \\
\\
NGC 4926 \\
   4.90 & 29 & 62.8 & $0.459 \pm 0.085$ \\
   7.35 & 35 & 141.4 & $0.248 \pm 0.042$ \\
  11.02 & 57 & 318.1 & $0.179 \pm 0.024$ \\
  16.53 & 59 & 571.1 & $0.103 \pm 0.013$ \\
  24.80 & 53 & 837.8 & $0.063 \pm 0.009$ \\
  37.20 & 668 & 2150.6 & $0.031 \pm 0.004$ \\
  55.80 & 81  & 5050.4 & $0.016 \pm 0.002$ \\
  83.70 & 107 & 7643.4 & $0.014 \pm 0.002$ \\
 125.56 &  8 & 1012.8 & $ 0.008 \pm 0.003$ \\
\\
IC 4051 \\
   3.67 & 101& 35.2 & $2.870 \pm 0.286$ \\
   5.51 & 134& 79.5 & $1.685 \pm 0.146$ \\
   8.27 & 215& 179.1 & $1.200 \pm 0.082$ \\
  12.40 & 313& 402.4 & $0.778 \pm 0.044$ \\
  18.60 & 269& 620.9 & $0.433 \pm 0.026$ \\
  27.90 & 245 & 1099.0 & $0.223 \pm 0.014$ \\
  41.85 & 284 & 2820.8 & $0.101 \pm 0.006$ \\
  62.78 & 325 & 6063.7 & $0.054 \pm 0.003$ \\
  94.17 & 277& 6260.1 & $ 0.044 \pm 0.003$ \\
 141.25 &  8 &  246.8 & $ 0.032 \pm 0.012$ \\
\enddata
\label{tab:rad}
\end{deluxetable}

\begin{deluxetable}{rrrrrrr}
\tablewidth{0pt}
\tablecaption{GCS Luminosity Function Data }
\tablehead{
\colhead{$V(min)$} & \colhead{$V(max)$} & \colhead{N4874} & \colhead{N4889} 
& \colhead{N4926} & \colhead{I4051} & \colhead{Bkgd} \\
}
\startdata
\\
   23.00 & 23.25 &   5  &  1  &    0 &   3 &  0  \\
   23.25 & 23.50 &  14  &  0  &    0 &   4 &  2  \\
   23.50 & 23.75 &  16  &  4  &    0 &  12 &  2  \\
   23.75 & 24.00 &  27  &  6  &    1 &  12 &  0  \\
   24.00 & 24.25 &  58  &  7  &    2 &  17 &  1  \\
   24.25 & 24.50 &  77  & 19  &    6 &  29 &  4  \\
   24.50 & 24.75 &  98  & 39  &    4 &  38 &  5  \\
   24.75 & 25.00 & 118  & 65  &    7 &  59 &  5  \\
   25.00 & 25.25 & 139  & 87  &    9 &  87 &  6  \\
   25.25 & 25.50 & 180  & 119 &   24 & 126 & 13  \\
   25.50 & 25.75 & 216  & 141 &   29 & 133 & 15  \\
   25.75 & 26.00 & 252 &  252 &   36 & 156 & 34  \\
   26.00 & 26.25 & 294 &  262 &   56 & 227 & 24  \\
   26.25 & 26.50 & 366 &  315 &   62 & 268 & 61  \\
   26.50 & 26.75 & 401 &  404 &   93 & 305 & 55  \\
   26.75 & 27.00 & 493  & 475 &   95 & 372 & 89  \\
   27.00 & 27.25 & 552  & 588 &  121 & 391 & 128 \\
   27.25 & 27.50 & 544  & 661 &  190 & 464 & 157 \\
   27.50 & 27.75 & 371  & 713 &  228 & 411 & 188 \\
   27.75 & 28.00 & 206 &  743 &  254 & 398 & 201 \\
   28.00 & 28.25 &  -  &  649 &  256 & 385 & 282 \\
   28.25 & 28.50 &  -  &   -  &  317 & 399 & 344 \\
\\
Area     &      & 4.80 & 4.44 & 2.60 & 3.34 & 3.99  \\
\enddata
\label{tab:lf}
\end{deluxetable}

\begin{deluxetable}{lcccc}
\tablewidth{0pt}
\tablecaption{Gaussian Model Fits to $(V-I)$ Color Distributions }
\tablehead{
\colhead{Galaxy} & \colhead{Mean $\mu$} & \colhead{Dispersion $\sigma$}
& \colhead{Proportion} & \colhead{Comment} \\
}
\startdata
 NGC 4874 & $1.106 \pm 0.006$ & $0.134 \pm 0.004$ & 1.00 & Unimodal\\ 
  & $1.001 \pm 0.042$ & $0.123 \pm 0.024$ & 0.30 & Bimodal \\ 
  & $1.147 \pm 0.018$ & $0.116 \pm 0.012$ & 0.70 & '' \\ 
 NGC 4889  & $1.096 \pm 0.007$ & $0.132 \pm 0.005$ & 1.00 & Unimodal \\ 
  & $0.978 \pm 0.016$ & $0.080 \pm 0.010$ & 0.30 & Bimodal \\ 
  & $1.147 \pm 0.011$ & $0.117 \pm 0.009$ & 0.70 & ''\\ 
 IC 4051 & $1.093 \pm 0.009$ & $0.157 \pm 0.006$ & 1.00 & Unimodal \\ 
  & $0.958 \pm 0.031$ & $0.123 \pm 0.020$ & 0.30 & Bimodal \\ 
  & $1.150 \pm 0.015$ & $0.132 \pm 0.013$ & 0.70 & ''\\ 
 NGC 4881 & $0.788 \pm 0.016$ & $0.093 \pm 0.012$ & 1.00 & Unimodal \\ 
 NGC 4926 & $1.191 \pm 0.016$ & $0.189 \pm 0.013$ & 1.00 & Unimodal \\ 
\enddata
\label{tab:cfit}
\end{deluxetable}

\clearpage

\clearpage

\clearpage

\clearpage

\clearpage

\clearpage

\clearpage

\clearpage

\clearpage

\clearpage

\clearpage

\clearpage

\clearpage

\clearpage

\clearpage

\end{document}